# Toward the mechanics of fractal materials: mechanics of continuum with fractal metric

Alexander S. Balankin

*Grupo "Mecánica Fractal", ESIME-Zacatenco, Instituto Politécnico Nacional, México D.F. 07738, Mexico*

This paper is devoted to the mechanics of fractal materials. A continuum framework accounting for the topological and metric properties of fractal domains in heterogeneous media is developed. The kinematics of deformations is elucidated and the symmetry of the Cauchy stress tensor is established. The mapping of mechanical problems for fractal materials into the corresponding problems for the fractal continuum is discussed. Stress and strain distributions in elastic fractal bars are analyzed. Some features of acoustic wave propagation and localization in scale-invariant media are briefly discussed. The effect of fractal correlations in the material microstructure on the crack mechanics is revealed. It is shown that the fractal nature of heterogeneity can either delay or assist the crack initiation and propagation, depending on the interplay between metric and topological properties of the fractal domain.

**I. Introduction**

Most natural and engineering materials are inherently heterogeneous [1]. The concept of continuum introduces an approximation of real medium by a region of Euclidean space filled by matter with continuous properties, where the term "continuous" refers that the material properties averaged on the length and time scales of interest vary smoothly, except possibly for a finite number of discontinuities. Accordingly, the continuum mechanics comes into play when one examines what is going on inside a body in a smoothed picture that does not go into details about the forces and motions of the sub-scale constituents. In this regard, traditional homogenization methods provide an efficient way to model the mechanical behavior of heterogeneous materials if the length scales are decoupled and the material microstructure has certain translational symmetry [2]. However, (micro-)structures of real heterogeneous materials frequently possess formidably complicated architecture exhibiting statistical scale invariance over many length scales [3,4]. Examples range from gels [5], polymers [6], and biological materials [7] to rocks [8], soils [9], and carbonate reservoirs [10]. For such materials the classical homogenization methods are inapplicable, because heterogeneities play an important role on almost all scales. This is reflected in the material response to external forces [11,12,13,14,15,16]. Hence, mechanics of scale-invariant materials is of tremendous importance for both fundamental and technological interest.

In this background, the fractal geometry offers helpful scaling concepts to characterize the scale invariant domains in heterogeneous materials [17,18,19,20]. These include the scale-invariant spatial and size distributions of solid phases and/or defects (e.g. pores or fractures), long-range correlations in the mass (or pore) density distribution, fractal geometry of fracture, pore, and crumpling networks, among others [3-21]. A key advantage of the fractal approach is the possibility to store the data relating to all scales of observation using a relatively small number of parameters that define a structure of greater complexity and rich geometry [22]. Unfortunately, the functions defined on fractals are essentially non-differentiable in the conventional sense [23]. This demands the development of novel tools to deal with fractal materials within a continuum framework.

One of them is the concept of non-local fractional derivative [24,25,26,27]. However, the use of non-local fractional calculus implies (reflects) the existence of long-term spatio-temporal memory in the medium [28]. Hence, the non-local fractional calculus may be suitable in cases when the physical nature of this memory is clear, but not in others. In the last cases, one wants to describe the kinematics of deformable fractal media using the local differential operators, despite the existence of long-range correlations in the material structure [29,30,31,32]. In this context, the introduction of differentiable analytic envelopes of non-analytic fractal functions [29] involves, at least implicitly, a continuum approximation of fractal medium. Explicitly, the notion of local fractal continuum was put forward by Tarasov [33]. Further, the fractal continuum approach was employed in Refs. [25,27,31,34,35,36,37,38,39,40,41,42]. The fractal continuum approximation allows us to define the macro properties of heterogeneous materials and express them through the structural parameters. This permits the use of well developed mathematical tools for solving mechanical engineering problems within a continuum framework.

However, some fundamental questions regarding to the definition of fractal continuum still remain under debate (see Refs. [43,44]).

In the present paper, we put forward a fractal continuum approach accounting for the topological, as well as the metric properties of fractal materials. The paper is organized as follows. Sec. II is devoted to scaling features of fractally heterogeneous materials. Dimension numbers characterizing the scale-invariance, topology, connectivity, and dynamics of fractal medium are outlined. The fractal-continuum homogenization of fractal media is discussed in Sec. III. In this context, the metric, norm, and measure accounting for the scaling properties of heterogeneous materials are introduced. Consequently, the local derivative and generalized Laplacian in the fractal continuum are defied. Sec. IV is devoted to the mechanics of fractal continua. The kinematics of fractal continuum deformations is developed. The Jacobian of transformations is established. Equations of the momentum conservation are derived. Forces and stresses in the fractal continuum are defined. Constitutive laws for fractal continuum are discussed. The mapping of mechanical problems for fractal materials into problems for fractal continua is elucidated in Sec. V. Some specific problems related to mechanics of fractal materials are briefly discussed. Some relevant conclusions are highlighted in Sec. VI.

## II. Scaling features of fractally heterogeneous materials

Generally, a heterogeneous material consists of domains of different materials, or the same material in different phases. Although, in mathematics, fractals can be defined without any reference to the embedding space [45], the natural and engineering materials reside in the three-dimensional space $E^3$ and occupy a well defined volume $V_3 \in E^3$. Accordingly, a fractal material necessary consists of fractal and non-fractal domains. For example, if the matrix of porous material is a fractal, the porous space cannot be a fractal and controversially, if the pore space is a fractal, the matrix should be a non-fractal [46]. Furthermore, in both cases the interface between solid matrix and pore space can also be a fractal [47].

The scaling properties of a fractal domain can be characterized by a set of fractional dimensionalities [19,48,49,50]. Most definitions of the fractional dimension numbers are based on the paradigm of domain covering by balls (cubes, tubes, etc.) of some size $\varepsilon$, or at most $\varepsilon$ [23,51,52]. In mathematics these covers are considered in the limit $\varepsilon \rightarrow 0$ and not necessary associated with the scale invariance of studied patterns. At the same time, it was noted that in many cases the number of n-dimensional covers need to cover a fractal of characteristic linear size $L$ scales as

$$N(L/\varepsilon) \propto \left(L/\varepsilon\right)^D, \tag{1}$$

where $\varepsilon$ is the length resolution scale, $d < D < n$ is the fractal (metric or box-counting) dimension, and $d$ is the topological dimension of the fractal pattern, while $n$ is the dimension of the embedding Euclidean space $E^n$ [17,18,53,54]. Examples include most classical fractals, such as the Cantor dusts (see Fig. 1), Koch curves, Sierpinski gaskets and carpets (see Fig. 2), Menger sponge (see Fig. 3), and percolation clusters [49,50], among others [17-23].

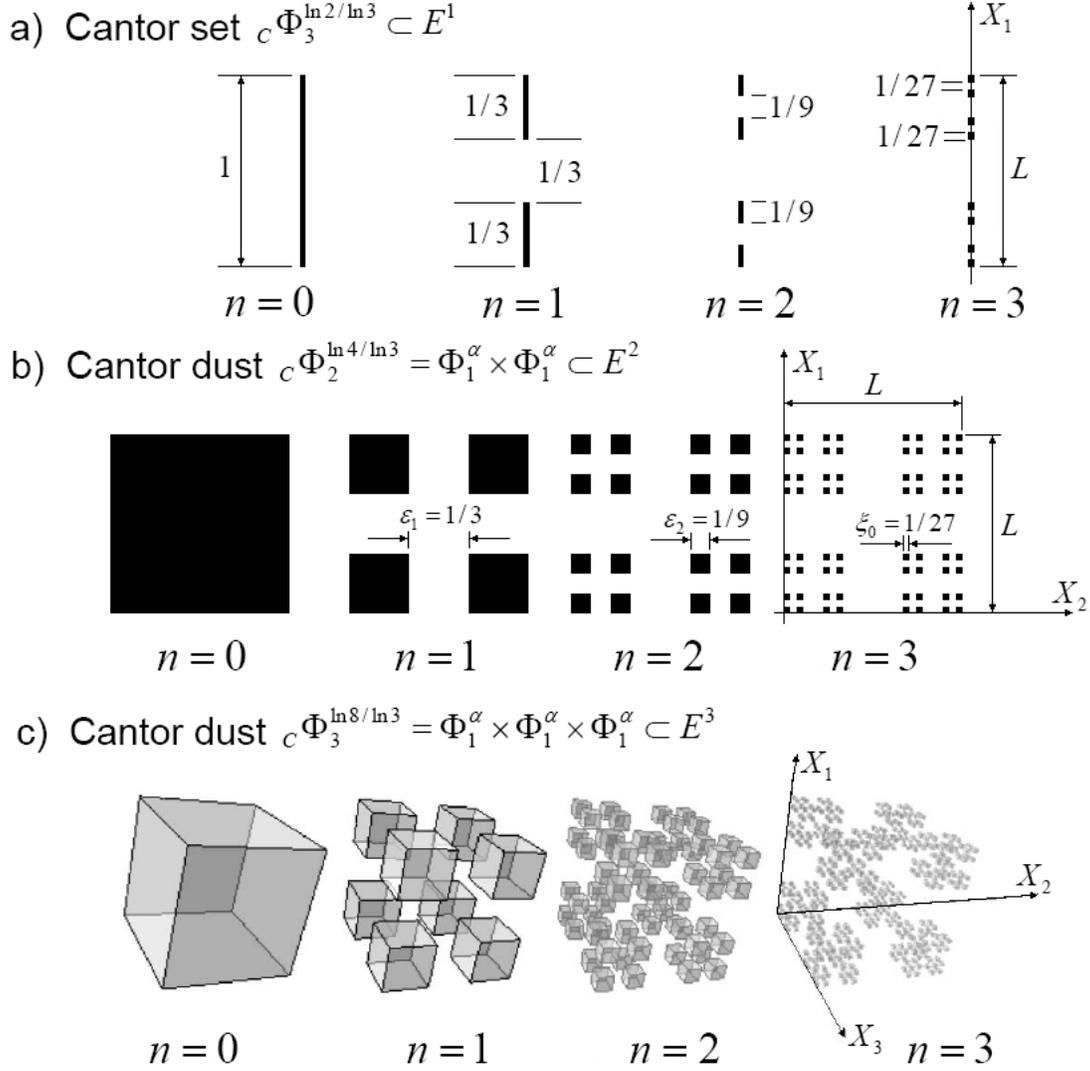

**Figure 1.** Iterative construction of Cantor dusts $_c\Phi_n^D = \prod_i^n \Phi_1^{\alpha_i}$ embedded into: (a) $E^1$ ($D = \ln 2/\ln 3$), (b) $E^2$ ($D = \ln 4/\ln 3$), and (c) $E^3$ ($D = \ln 8/\ln 3$). Notice that the topological dimension of any Cantor dust is $d = 0$, whereas the intrinsic fractal dimension and spectral dimensions are equal to the dimension of the embedding Euclidean space $E^n$, that is $0 = d < D < d_\ell = d_s = n$ [64].

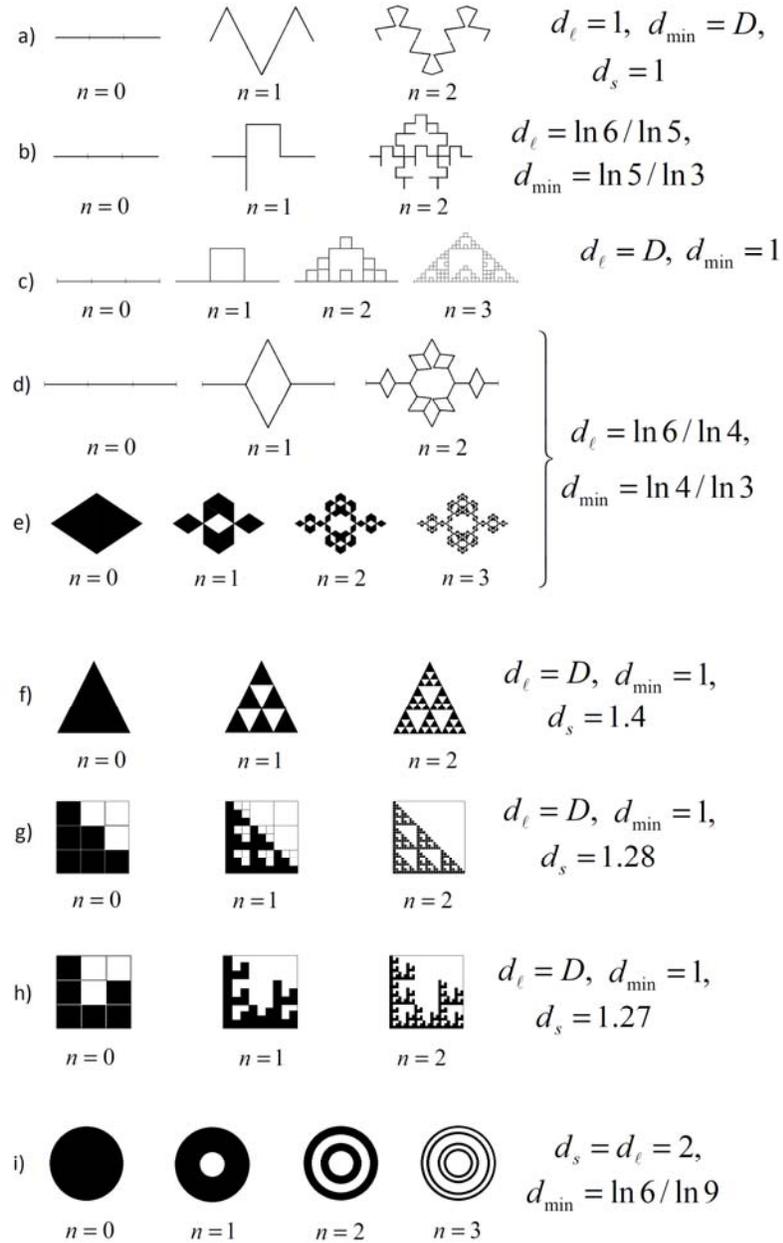

**Figure 2.** Iterative constructions of 9 fractals of the same topological $d = 1$ and fractal ($D = \ln 6 / \ln 3 < n = 2$) dimensions, but having different topological and connectivity properties: (a-e) Koch curves; (f) Sierpinski gasket; (g,h) Sierpinski carpets and (i) Cantor circles. The values of spectral dimensions are taken from Refs. [55].

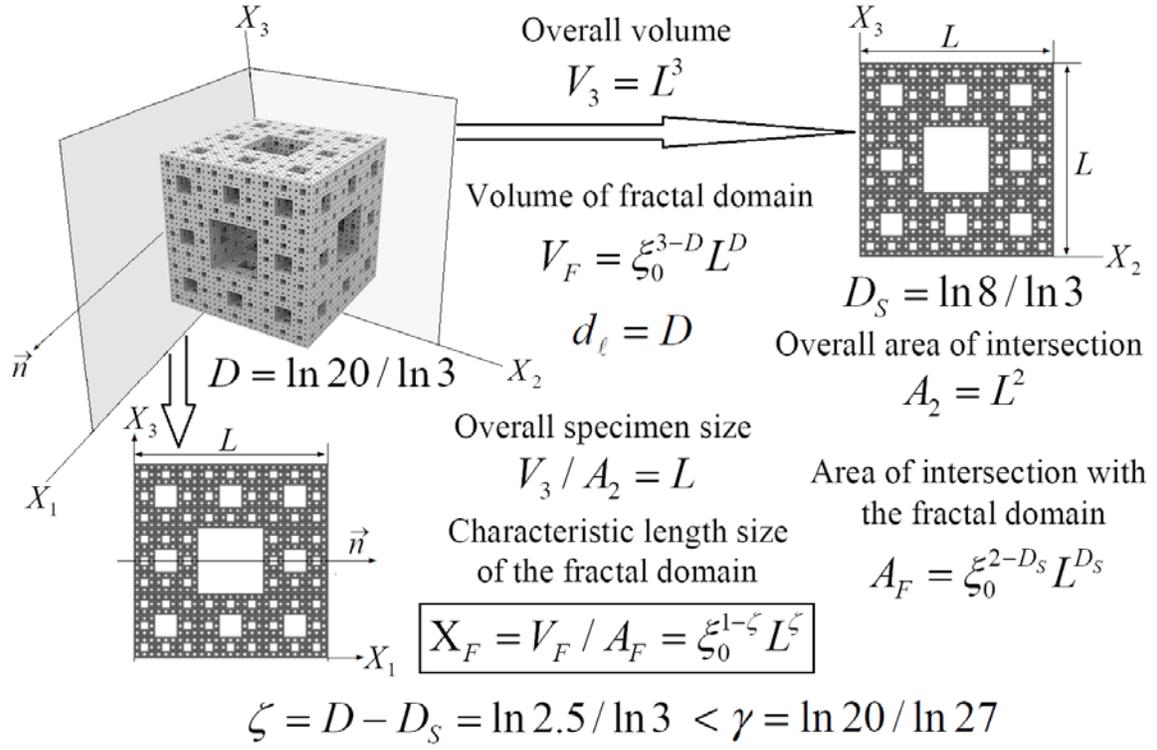

**Figure 3.** Definitions of the intersection area and the characteristic length scale in the direction normal to the intersection between the Menger sponge ($D = \ln 20 / \ln 3$) and the Cartesian plane in $E^3$. The topological dimension of Menger sponge is $d = 1$, whereas the intrinsic fractal dimension is $d_\ell = D < n = 3$ ($d_{min} = 1$) and the spectral dimension is $d_s \approx 2.5$. The intersection of Menger sponge with plane is the Sierpinski carpet of the fractal dimension $D_S = \ln 8 / \ln 3 \approx 1.893$, while the intrinsic fractal dimension is $d_\ell^{SC} = D_S$, and so the co-dimension is $\zeta_i = D - D_S = \ln 2.5 / \ln 3 \approx 0.83 < \gamma = d_\ell / 3 \approx 0.91$.

It is precisely the power-law behavior (1) that allows us to use the powerful tools of fractal geometry for deal with fractal materials exhibiting statistical scale invariance over a bounded interval of length scale

$$\xi_0 < \varepsilon \le L < \xi_C,$$ (2)

where $\xi_0$ and $\xi_C$ are the lower and upper cut-offs of physical origin [4,56], whereas the "physical" fractal dimension $D$ can be associated with some kind of box-counting quasi-measure [57]. So, strictly speaking, fractal materials exhibit pre-fractal, rather than true fractal features.

The mass of a pre-fractal domain $\Phi_3^D$ in $E^3$ scales with its characteristic linear size as

$$M = \rho_0 \xi_0^3 \left( \frac{L}{\xi_0} \right)^D \text{ for } \xi_0 < L < \xi_C, \tag{3}$$

whereas

$$M = \rho_0 L^3, \text{ if } L < \xi_0, \text{ and } M = \rho_m L^3, \text{ if } L >> \xi_C,$$

where $\xi_0$ is the characteristic size of elemental Euclidean components of the mass density $\rho_0$ from which the pre-fractal domain is made up (see Fig. 1) and $\rho_m$ is the overall density of the fractal material, e.g. $\rho_m = \rho_0 (1 - \phi)$, while $\phi$ is the total porosity [38]. Accordingly, the physical fractal dimension $D$ is experimentally measurable from the power law behavior of extensive (mass, number of structural components, surface area, *etc*.) or intensive (mass density, two-point correlation function, *etc*.) properties of the studied fractal material.

The fractal dimension $D$ characterizes how the properties of a fractal domain change with the body size $L$ within the interval (2). However, the knowledge of $D$ alone is insufficient to characterize the scaling properties of the fractal domain. In fact, fractals of the same mass (metric) dimension $D$ can have very different topology, connectivity, and dynamic properties (see Fig. 2). Hence, to define the scaling properties of fractal material in a non-ambiguous way, one needs to employ some more dimension numbers.

Specifically, the connectivity and topological properties of a fractal pattern are characterized by the so-called intrinsic fractal dimension $d_\ell \leq n$ [48,58], also termed as the connectivity dimension [59], the chemical dimension [49,50,60,61], and the spreading dimension [54]. In physics, this dimensionality is defined via the scaling relation

$$N(\ell) \propto \left(\ell / \varepsilon\right)^{d_\ell},\qquad(4)$$

where $N(\ell)$ is the number of covering elements of size $\ell \propto R^{d_{\min}}$, while $d_{\min} = D / d_\ell$ is the fractal dimension of the shortest (chemical) path $\ell$ connecting two randomly chosen sites in the (pre-)fractal domain and $R$ is the Euclidean distance between these sites [48-50]. Therefore, $d_\ell$ quantifies how the "elementary" structural units are "glued" together to form the entire (pre-)fractal object in the embedding Euclidean space $E^n$ [60]. Furthermore, it is easy to see that $d_\ell$ defiled by Eq. (4) is intimately linked with the intrinsic fractal dimension introduced in [45] and so it tells us "how many directions" the observer feels in the fractal space by making static measurements [62]. Hence, $d_\ell \geq d$ determines the minimal number of independent coordinates needs to unambiguously define the point position in the fractal medium, in the same way as the topological dimension $d \leq n$ determines the number of orthogonal coordinates in the Euclidean manifold (e.g. $d = 3$ for a box, $d = 2$ for a smooth surface, and $d = 1$ for a differentiable curve). Notice that, per definition, the intrinsic fractal dimension of a fractal is always greater or equal to one and $D \geq d_{\min}$ (see Fig. 2), whereas the fractal dimension of the shortest path can be either $d_{\min} \geq 1$, if the fractal is path-connected (see Figs. 2 a-h, 3), or $d_{\min} < 1$, as this is in the case of totally discontinuous fractals, such as the Cantor dust (see Fig. 1), and fractals which are discontinuous along some Cartesian directions (see Fig. 4).

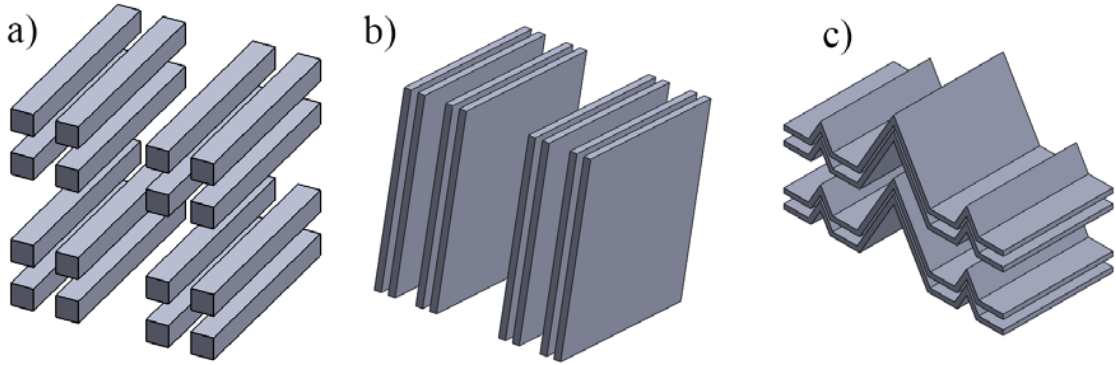

**Figure 4.** The second iterations of: (a, b) Cartesian product (7) with: (a) $\alpha_1 = \alpha_2 = \ln 2 / \ln 3$ and $\alpha_3 = 1$; (b) $\alpha_1 = \ln 2 / \ln 3$ and $\alpha_2 = \alpha_3 = 1$; and (c) Cartesian product (9) with $D_I = \ln 4 / \ln 3$, $\alpha_1 = \ln 2 / \ln 3$ and $\alpha_1 = \ln 2 / \ln 4 = 0.5$.

Another important topological characteristics of fractal materials are the fractal dimensions $D_S^{(i)}$ of intersections between the fractal domain $\Phi_3^D \subset E^3$ and two-dimensional Cartesian planes in $E^3$ (index $i = 1,2,3$ denotes the Cartesian plane orthogonal to $i$-axis) [38-41]. Accordingly, there were many attempts to establish a relation between $D_S^{(i)}$ and $D$. In this way it was found that some kinds of mathematical and physical fractals obey the Mandelbrot rule of thumb [17,63]:

$$D_S^{(i)} = D - 1,$$  (5)

for example, percolation clusters [64] and some kinds of porous soils [18]. On the other hand, in the case of fractals which can be treated as the Cartesian product

$$_C\Phi_3^D = \Phi_1^{\alpha_1} \times \Phi_1^{\alpha_2} \times \Phi_1^{\alpha_3} \subset E^3$$  (6)

of three manifolds $\Phi_1^{\alpha_i} \subset E^1$ with the fractal dimensions $\alpha_i \leq 1$, where at least one $\alpha_i < 1$ (see Figs. 1 and 4a,b), the fractal dimension of intersection with the Cartesian plane is equal to

$$D_S^{(i)} = D - \alpha_i = \alpha_j + \alpha_k,$$  (7)

where $j \neq k$ [23]. The intrinsic fractal dimension of the Cartesian product (6) is $d_\ell = 3$ [65], whereas its fractal (mass, box-counting) dimension is equal to

$$D = d_{\min}^{(1)} + d_{\min}^{(2)} + d_{\min}^{(3)},$$  (8)

where $d_{\min}^{(i)} = \alpha_i \leq 1$ are the fractal dimensions of minimum path along the Cartesian axes. Notice that the inequality $d_{\min}^{(i)} < 1$ means that the fractal domain $_C\Phi_3^D \subset E^3$ is discontinuous along $i$-axis. Specifically, equalities (7) and (8) hold for totally discontinuous Cantor dusts $_C\Phi_3^D$ with all $\alpha_i < 1$ (see Fig. 1c) and for fractals which can be treated either as the Cartesian product of the Cantor dust $_C\Phi_2^D$ with the Euclidean line (see Fig. 4a), or the Cantor set $_C\Phi_1^\alpha$ with the Euclidean plane (see Fig. 4b). In this regard, although the inequality $d_\ell = 3 > D$ implies that the fractal is discontinuous in $E^3$, it doesn't necessarily mean that the fractal can be treated as the Cartesian product (6) obeying the equality (8). In fact, one can construct the Cartesian product

$$_3\Phi_3^D = \Phi_2^{D_I} \times \Phi_1^{\alpha_2} \times \Phi_1^{\alpha_3} \subset E^3 \qquad (9)$$

of the fractal with $1 < D_I < 2$ and two fractals with $\alpha_2, \alpha_3 \le 1$, such that $\alpha_2 + \alpha_3 < 2$ and

$$D = D_I + \alpha_2 + \alpha_3 < d_\ell = 3 , \qquad (10)$$

but the equality (8) does not hold. For example, the fractal shown in Fig. 4c has $d_{\min}^{(1)} = \ln 2 / \ln 3$, $d_{\min}^{(2)} = 1/2$, $d_{\min}^{(3)} = 1$, and so $d_{\min}^{(1)} + d_{\min}^{(2)} + d_{\min}^{(3)} < D < d_\ell = 3$, whereas

$$D_S^{(1)} = D - D_I = \alpha_2 + \alpha_3, \ D_S^{(2)} = D - \alpha_2 = D_I + \alpha_3, \text{ and } D_S^{(3)} = D - \alpha_3 = D_I + \alpha_2 . \quad (11)$$

Furthermore, for fractals with $2 \le d_\ell < 3$ the fractal dimensions of intersections are generally independent of $D$ and can take any values in the ranges $0 < D_S^{(i)} \le 2$, even if $D_S^{(i)} = D_S$ for any intersection [23], except of the special case of fractals obeying the Mandelbrot rule of thumb (5).

The fractal dimension $D_S^{(i)}$ characterizes how the "effective" area $A_\partial^{(i)}$ of intersection between the (pre-)fractal and two-dimensional Cartesian plane $(x_j, x_k)$ scales with its overall size $L$. Specifically, in the case of pre-fractal domains with $d_\ell \geq 2$, the area of this intersection scales as $A_\partial^{(i)} = \xi_0^2 (L/\xi_0)^{D_S^{(i)}}$ [51], while $L$ varies within the interval (2), and so Eq. (3) can be re-written in the form

$$M = \rho_0 X^{(i)} A_\partial^{(i)} = \rho_0 \xi_0^3 (L/\xi_0)^D , \tag{12}$$

where $X^{(i)}(L) = \xi_0 (L/\xi_0)^{\zeta_i}$ and

$$\zeta_i = D - D_S^{(i)} > 0 \tag{13}$$

is the co-dimension of intersection in $\Phi_3^D$ (see Fig. 3). In this respect, it should be pointed out that, generally, $\zeta_i \leq 1$, but $\sum_i^3 \zeta_i = 3D - \sum_i^3 D_S^{(i)} \neq D$, except of the case (6). Specifically, in fractals obeying the Mandelbrot rule of thumb (5) all co-dimension are equal to one (that is $\zeta_i \equiv 1$) and so $\sum_i^3 \zeta_i = 3 > D$, whereas for the fractal shown in Fig 4c it is a straightforward matter to calculate that $\sum_i^3 \zeta_i < D < d_\ell = 3$. Even so, the mass scaling of this fractal can be presented in the form of Eq. (12) as $M \propto L^{\zeta_i} A_\partial^{(i)} \propto L^{\zeta_i} (\sqrt{L \times L})^{D_S^{(i)}} \propto L^D$, where $\zeta_i$ and $D_S^{(i)}$ are defined by Eqs. (13) and (11), respectively. Furthermore, although the Menger sponge with $D = \ln 20 / \ln 3 = d_\ell < 3$ (see Fig. 3), cannot be viewed as the Cartesian product, the scaling behavior of the Menger sponge mass also can be presented in the form of Eq. (13) as follows

$$M \propto L^{\zeta_i} A_0^{(i)}(L) = L^\zeta \left(L \times L\right)^{D_S / 2} = L^D, \qquad (14)$$

where $D_S^{(i)} = \ln 8 / \ln 3$ is the fractal dimension of the Sierpinski carpet (see Fig. 3), such that all co-dimensions $\zeta_i$ defined by Eq. (13) are equal to $\zeta = \ln(2.5) / \ln 3$, and so $\sum_i^3 \zeta_i = \ln(15.625) / \ln 3 < D$ and $\sum_{j \neq i}^2 \zeta_j = \ln(6.25) / \ln 3 < D_S^{(i)}$.

Dimension numbers defined above can be used to distinguish between fractal domains having the same fractal (mass) dimension $D$ but different connectivity and topological properties (see Figs. 1-4). However, to describe all scaling properties of a specific fractal domain in an unambiguous way, one may need to use some more independent scaling exponents (see Refs. [19,48-50,66,67,68,69,70]). Specifically, dynamical properties of a fractal domain are governed by its spectral dimension $d_s$ defined via the scaling relation $\Omega(\omega) \propto \omega^{d_s - 1}$, where $\Omega(\omega)$ is the density of fractal vibration modes with frequency $\omega$ [71,72]. The spectral dimension is therefore fundamental for any diffusive process, such as the random walk on the fractal domain [73]. In practice, the number of independent scaling exponents which should be employed to characterize a specific fractal domain depends on the problem under consideration. For example, if one is only interested in the overall porosity of fractal material, the knowledge of the fractal (mass) dimension of the pore space may be sufficient. However, if we are interesting in the fluid flow through a fractal material, we need to know the fractal and intrinsic fractal dimensions of the pore network backbone [74], as well as the fractal dimensions of its intersections with the Cartesian planes [39] and, in some cases, the fractal dimension of the pore-solid interface [75]. Furthermore, to describe diffusion processes in the pore network, one also needs to know its spectral dimension [72]. The spectral dimension also governs the stress and strain relaxation in fractal materials [76]. So, to describe mechanical properties of a fractal, e.g. the percolation cluster, it may be necessary to define its elastic (rigid) backbone [77] and to determine the corresponding fractional dimensionalities. In this paper we limiting ourselves to the case of fractal domains for which the backbone and elastic backbone coincide with the whole fractal.

## III. Fractal continuum homogenization of fractal media

Fractal domain $\Phi_3^D \subset E^3$ of the fractal dimension $d < D < 3$ cannot continuously fill the embedding Euclidean space and so its properties are almost everywhere discontinuous in $E^3$ [23]. Despite of this, one can suppose a virtual fractal continuum $\Phi_D^3 \subset E^3$ having the topological dimension $d = 3 > D$, the properties of which (density, displacements, etc.) are defined as analytic envelopes of non-analytic functions in the fractal material under study. The constitutive condition

$$d = 3 > D \qquad (15)$$

can be fulfilled if the density of admissible states in $\Phi_D^3 \subset E^3$ is scale dependent [25,39]. The scale dependence of density of admissible states can be either introduced as the constitutive assumption for a fractal continuum with the Euclidean metric (see Refs. [25,33-36]), or can be a consequence of the postulated fractal metric in $\Phi_D^3 \subset E^3$ (see Refs. [38-41]). In this way, the concept of fractal continuum provides an efficient homogenization method for fractal media.

In the present work, we define the fractal continuum $\Phi_D^3 \subset E^3$ as a three-dimensional region of the embedding Euclidean space $E^3$ filled with continuous matter (such that $d = 3$) and endowed with appropriate fractional measure, metric, and norm, as well as with a set of rules for integro-differential calculus and a proper Laplacian, accounting for the metric, connectivity, and topological properties of the fractal domain $\Phi_3^D \subset E^3$ in the material under study.

### 3.1. The metric, norm, and measure in fractal continuum

The (quasi-)measure in $\Phi_D^3 \subset E^3$ is defined in such a way that the mass of any cubic (or spherical) region $W(L) \subset \Phi_D^3$ obeys the power-law behavior (3) with the fractal dimension $D$ equal to the mass fractal dimension of the fractal domain $\Phi_3^D \subset E^3$ under study [25]. However, this requirement does not lead to unique definitions of measure and metric in $\Phi_D^3 \subset E^3$. Therefore, additional assumptions are required to develop a fractal continuum model accounting for the essential features of a specific problem for the studied fractal medium. For example, to define the fractal measure in the fractal continuum, Tarasov [33] has suggested to use the Reitz potential, whereas the metric in the fractal continuum is assumed to be the conventional Euclidean metric. Notice that the use of a specific fractional integral is equivalent to the constitutive assumption about the functional form of scale dependent density of states in the fractal continuum model. In this way, to account for anisotropy of fractal materials one can also exploit the Cartesian product measure allied with the Cartesian decomposition (6) together with a suitable multiple fractional integral [25,35,78].

Accordingly, let us first consider the totally discontinuous Cantor dust in $E^3$ (see Fig. 1c). The intrinsic fractal dimension of Cantor dust is $d_\ell = 3$ [64] and so the position of any point $A \in _C \Phi_3^D \subset E^3$ can be specified by three Cartesian coordinates ($x_{ai}$). Consequently, one can define the distance $\Delta(A, B)$ between two points $A, B \in _C \Phi_3^D \subset E^3$ as the Euclidean norm of the difference between two vectors $\vec{a} = (x_{a1}, x_{a2}, x_{a3}) \in E^3$ and $\vec{b} = (x_{b1}, x_{b2}, x_{b3}) \in E^3$, where $x_{ai}$ and $x_{ai}$ are equal to the Euclidean lengths of vector projections on the Cartesian axes in $E^3$. Therefore, one can construct the corresponding fractal continuum

$$_C \Phi_D^3 = \Phi_{\alpha_1}^1 \times \Phi_{\alpha_2}^1 \times \Phi_{\alpha_3}^1 \subset E^3 \qquad (16)$$

endowed with the Euclidean norm and metric, whereas the fractal nature of Cartesian product (6) is accounted for by the introduction of scale dependent density of admissible states in the fractal continuum (16), as it was suggested in Refs. [33-37].

Alternatively, the "effective" lengths of vector projections in the fractal domain $_C\Phi_3^D \subset E^3$ can be defined as follows

$$|\chi_{ai}| = \xi_0^{1-\alpha_i}|x_{ai}|^{\alpha_i} \text{ and } |\chi_{bi}| = \xi_0^{1-\alpha_i}|x_{bi}|^{\alpha_i}, \qquad (17)$$

as it is shown in <span style="color:red">Fig. 5</span>. Consequently, one can define the orthogonal "fractal" coordinates $\chi_i$ of points $A$ and $B$ in the fractal continuum (16) as

$$\chi_i = sign(x_i)\xi_0^{1-\alpha_i}|x_{ai}|^{\alpha_i}, \qquad (18)$$

where the scaling exponents are equal to the fractal dimensions of the minimum path $\alpha_i = d_{\min}^{(i)} \leq 1$ in the Cantor sets $\Phi_1^{\alpha_i} \subset E^1$ along the Cartesian axes in $E^3$ (see <span style="color:blue">Fig. 5</span>). So, the distance between projections of two points $A, B \in_C \Phi_D^3 \subset E^3$ on the $\chi_i$-axis in the fractal continuum can be defined as

$$\Delta_i(A,B) = |\chi_{ai} - \chi_{bi}| = \xi_0^{1-\alpha_i}\left| sign(x_{ai})|x_{ai}|^{\alpha_i} - sign(x_{bi})|x_{bi}|^{\alpha_i}\right|, \qquad (19)$$

where $\chi_{ai}, \chi_{bi} \in \Phi_{\alpha_i}^1 \subset E^1$ denote the components of orthogonal fractal coordinates in the fractal continuum model (16) of the fractal domain $_C\Phi_D^3 \subset E^3$ (see <span style="color:blue">Fig. 5</span>). In this way, the fractal continuum (16) can be also equipped with the norm defined as

$$\|A\| = \sqrt{\chi_{a1}^2 + \chi_{a2}^2 + \chi_{a3}^3} \; , \tag{20}$$

and so the distance $\Delta(A,B)$ in $_C\Phi_D^3 \subset E^3$ is equal to

$$\Delta(A,B) = \sqrt{\Delta_1^2 + \Delta_2^2 + \Delta_3^2} \; , \tag{21}$$

where the distances $\Delta_i$ along the Cartesian axes in $_C\Phi_D^3 \subset E^3$ are given by Eq. (19), as this is illustrated in Fig. 5. It is a straightforward matter to verify that the distance defined by Eq. (21) together with Eq. (19) satisfies all conventional criteria required of metrics (see Ref. [40]).

Furthermore, in the fractal continuum (16) the infinitesimal volume element can be decomposed as

$$dV_D = d\chi_1 d\chi_2 d\chi_3 = c_1^{(1)}(x_1)c_1^{(2)}(x_2)c_1^{(3)}(x_3)dx_1 dx_2 dx_3 = c_3(x_k)dV_3 \; , \tag{22}$$

where $d\chi_i$ and $dx_i$ are the infinitesimal length elements in $_C\Phi_D^3 \subset E^3$ and in $E^3$, respectively, while

$$c_3 = c_1^{(1)}c_1^{(2)}c_1^{(3)} \tag{23}$$

and $c_1^{(i)}(x_i)$ can be interpreted as the densities of admissible states in the fractal continuum (16) and along the Cartesian axes in $_C\Phi_D^3 \subset E^3$, respectively [38-41]. From Eq. (22) together with Eq. (18) immediately follows that

$$c_1^{(i)} = \alpha_i \varsigma_0^{1-\alpha_i} |x_i|^{\alpha_i - 1} \tag{24}$$

and so, if $\alpha_i = D/3$ the density of admissible states in $_C\Phi_D^3 \subset E^3$ takes the form $c_3(R) = c_1^{(1)} c_1^{(2)} c_1^{(3)} \propto R^{D-3}$, where $R = (x_1 x_2 x_3)^{1/3}$. In this regard, it is pertinent to note that, although fractal continuum models of the Cartesian product (6) with the Euclidean metric (see Refs. [33-37]) and with the fractal metric defined by Eqs. (19) and (21) both have the same density of admissible states given by Eqs. (23) and (24), the kinematics of these models is quite different due to the difference of metrics (see Ref. [44]).

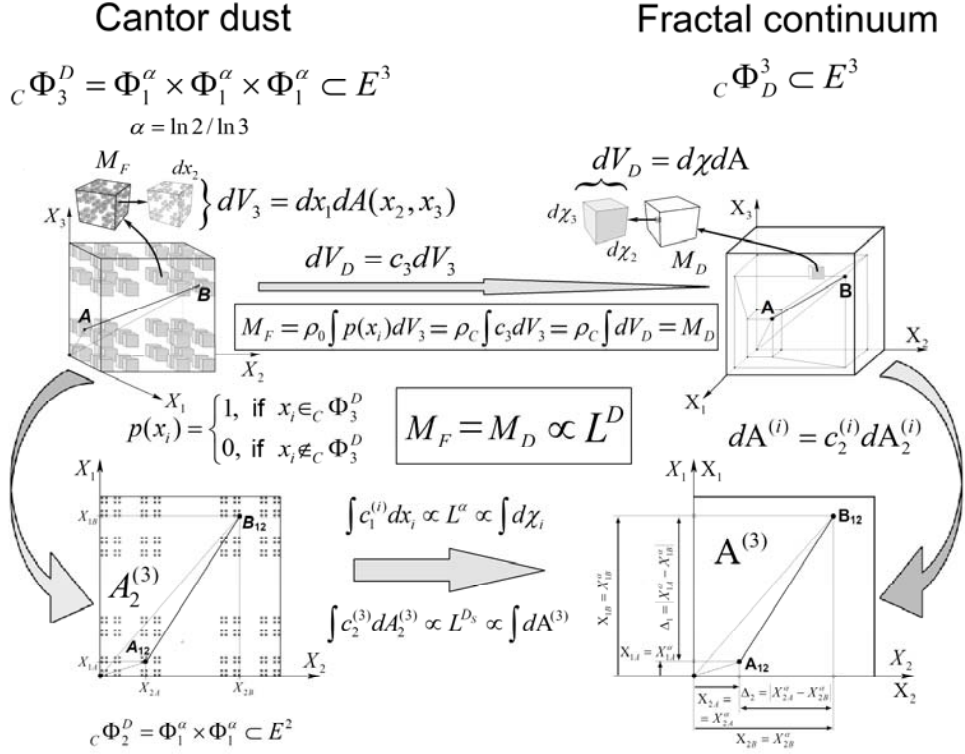

**Figure 5.** Mapping of the Cantor dust $_{C}\Phi_{3}^{D} \subset E^{3}$ into the fractal continuum $_{C}\Phi_{D}^{3} \subset E^{3}$ ($d_{\ell} = d_{s} = 3$, $\gamma = 1$, $\alpha_{i} = \ln 2/\ln 3$, $D = \ln 8/\ln 3$, and $D_{S} = \ln 4/\ln 3$) and geometric illustrations of the fractal norm (20), metric (19), (21), and measure (22).

In contrast to the Cartesian product (6), a path connected fractal domain $\Phi_{3}^{D} \subset E^{3}$ with $d_{\ell} < 3$ (see, for example, <span style="color:blue">Fig. 3</span>) cannot be represented in a three-dimensional Cartesian frame in a unique way, because the number of independent coordinates which can be defined in $\Phi_{3}^{D} \subset E^{3}$ is less than 3. Therefore, the infinitesimal volume element in the corresponding fractal continuum $\Phi_{D}^{3} \subset E^{3}$ cannot be decomposed as in Eq. (22).

However, taking into account the scaling relation (12), the infinitesimal volume element in the fractal continuum $\Phi_D^3 \subset E^3$ can be more generally decomposed as follows

$$dV_D = d\chi_i(x_i)dA_\vartheta^{(i)}(x_{j\neq i}) = c_1^{(i)}(x_i)c_2^{(i)}(x_{j\neq i})dx_i dA_2^{(i)} = c_3(x_k)dV_3 = c_3 dx_1 dx_2 dx_3, \quad (25)$$

where $dA_2^{(i)} = dx_j \cdot dx_k$ and $dA_\vartheta^{(i)}$ are the infinitesimal area elements on the intersection between $\Phi_D^3$ and two-dimensional plane normal to $i$-axis in $E^3$ and in $\Phi_D^3 \subset E^3$, respectively, while $c_2^{(i)}(x_{j\neq i})$ is the density of admissible states in the plane of this intersection (see Fig. 6). This allows us to define a pair of mutually orthogonal fractal coordinates ($\chi_i, A_\vartheta^{(i)}$) associated with the decomposition (25). Furthermore, from Eq. (25) together with the scaling relation (12) immediately follows that the transformation functions (densities of admissible states) in $\Phi_D^3 \subset E^3$ obey the following relationship:

$$c_3(x_i) = c_1^{(k)}(x_k)c_2^{(k)}(x_{j\neq k}), \quad (26)$$

but, for the path-connected fractals with the intrinsic fractal dimension $d_\ell < 3$,

$$c_2^{(k)}(x_i, x_j) \neq c_1^{(i)}(x_i)c_1^{(j)}(x_j),$$

because a choice of the coordinate pair ($\chi_i, A_\vartheta^{(i)}$) is not unique [79]. Consequently, in the corresponding fractal continuum the equality (23) does not hold, whereas the equality

(26) holds for any fractal continuum with $2 \leq d_\ell \leq 3$ [41]. Accordingly, to fulfill the constitutive requirement (12), the densities of admissible states in $\Phi_D^3 \subset E^3$ should obey the following scaling relations:

$$\int dV_3 c_3 \propto \xi_0^{3-D} L^D \text{ and } \int dA_2^{(i)} c_2^{(i)} \propto L^{D_s^{(i)}} \text{, while } \int d\chi_i = \int dx_i c_1^{(i)} \propto \xi_0^{1-\zeta_i} L^{\zeta_i}, \quad (27)$$

where the index $i$ denotes a Cartesian direction, the scaling exponent $\zeta_i$ is defined by Eq. (13), and the Lebesgue–Stieltjes integral is used. The third relationship of Eq. (27) immediately implies that

$$\chi_i = sign(x_i)\xi_0^{1-\zeta_i}\left|x_{ai}\right|^{\zeta_i} \quad (28)$$

and so

$$c_1^{(i)} = \zeta_i \xi_0^{1-\zeta_i}\left|x_i\right|^{\zeta_i-1}, \quad (29)$$

even when the equality (23) does not hold. Notice that relations (25)-(29) can be also used to construct the fractal continuum models of Cartesian products (6) and (9) which have the intrinsic fractal dimension $d_\ell = 3$. In the former case the equality (23) holds, whereas in the case of Eq. (9) it does not hold.

Furthermore, for some types of path-connected fractals the explicit functional forms of $c_2^{(i)}(x_{j \neq i})$ and so $c_3(x_k)$ can be also derived from the second relation of Eq. (27). Specifically, for a fractal continuum model of the Menger sponge (see Fig. 6) it is a straightforward matter to see that the densities of admissible states can be defined as follows

$$c_1^{(1)} = \zeta\xi_0^{1-\zeta}\left|x_1\right|^{\zeta-1}, \ c_2^{(1)}(x_2,x_3) = (D_S/2)^2\xi_0^{2-D_S}\left|x_2\right|^{(D_S/2)-1}\left|x_3\right|^{(D_S/2)-1},$$

and                                                                                                           (30)

$$c_3(x_i) = \zeta(D_S/2)^2\xi_0^{3-D}\left|x_1\right|^{D-D_S-1}\left(\sqrt{\left|x_2 x_3\right|}\right)^{D_S-2},$$

such that although the equality (23) does not hold, the scaling relation (14) and the equality (26) both do hold.

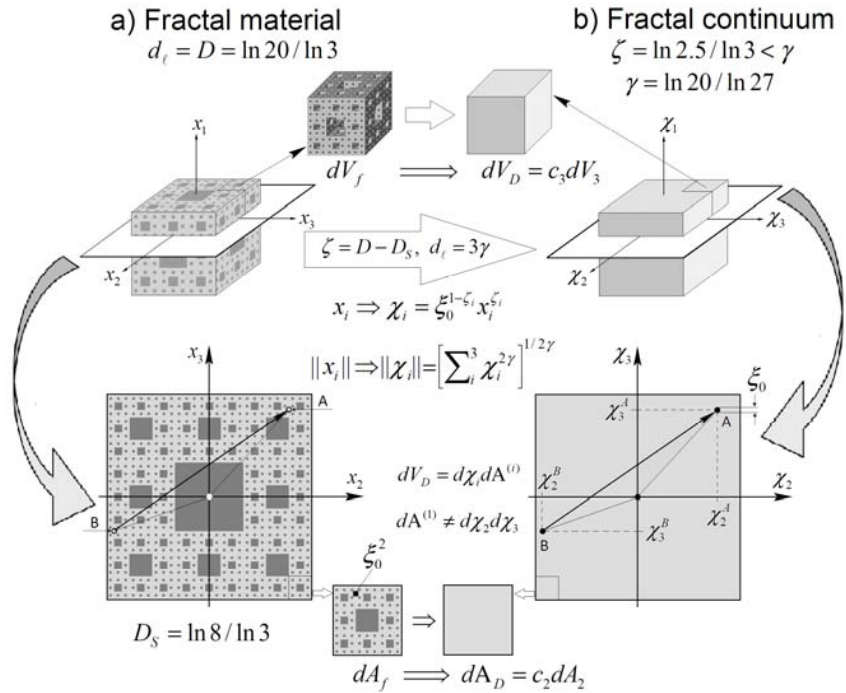

**Figure 6.** Mapping of the Menger sponge ($d_\ell = D = \ln 20/\ln 3 < 3$, $D_S = \ln 8/\ln 3$) into the fractal continuum $\Phi_D^3 \subset E^3$ with $\zeta = D - D_S = \ln 2.5/3$, $\gamma = d_\ell/3 \approx 0.91$, and geometric illustration of the fractal norm (32), metric (33), (34), and measure (25).

In contrast to a fractal the topological dimension of which is, per definition (see Ref. [17]), less or equal to its intrinsic fractal dimension (that is $d \leq d_\ell \leq n = 3$), the topological dimension of fractal continuum is equal to the dimension of the embedding Euclidean space per constitutive definition (15), that is $d_\ell \leq d = n = 3$. Hence, in the fractal continuum $\Phi_D^3 \subset E^3$ one always can define three independent fractal coordinates (28). In this background, the impossibility to define $c_2^{(i)}$ and $c_3$ in a unique way is an intrinsic feature of mapping from two mutually orthogonal fractal coordinates ($\chi_i$ and $A_\partial^{(i)}$) associated with decomposition (12) into three orthogonal fractal coordinates ($\chi_1, \chi_2, \chi_3$) in $\Phi_D^3 \subset E^3$ with $d = 3 > d_\ell$ [41]. So, in essence, this is a price one has to pay in order to deal with fractals which cannot be treated within the approach based on the Cartesian decomposition (6). Fortunately, this does not impose serious limitations, since we need not to know explicit functional forms of $c_3$ and $c_2^{(i)}$, as long as equality (26) holds [44].

Further, to account for the effect of fractal topology of the scale-invariant domain $\Phi_3^D \subset E^3$ with the intrinsic fractal dimension $2 \leq d_\ell < 3$ (e.g. the Menger sponge shown in Fig. 6) on the metric in the corresponding fractal continuum model $\Phi_D^3 \subset E^3$, let us consider a fractional dimensional space $F^\gamma$ in which the fractal domain $\Phi_3^D$ can be embedded. The axiomatic definition of a fractional dimensional space $F^\gamma$ was suggested by Stillinger [80] and further widely used in different areas of physics (see, for example,

Refs. [31,32,61,81,82,83] and references therein). In this context, Calcagni [61] has equipped the fractional dimensional space $F^\gamma$ with the fractional norm

$$\|A\| = \left[ \sum_i^3 x_{ai}^{2\gamma} \right]^{1/2\gamma} \tag{31}$$

accounting for the fractal topology of space $F^\gamma$ having the fractional dimension $d_f = 3\gamma \le 3$. Notice that for $1.5 \le d_f \le 3$ Eq. (31) mathematically coincides with the conventional definition of the p-norm with $1 \le p = 2\gamma \le 2$, which converts into the Euclidean norm in the limit $\gamma = 1$ ($d_f = 3$) and into the Manhattan norm in the limit $\gamma = 0.5$ ($d_f = 1.5$). So, Eq. (31) satisfies all conventional requirements of norm as long as $1.5 \le d_f \le 3$.

Following to Ref. [61], to account for the connectivity and topology of the fractal domain $\Phi_3^D \subset F^\gamma \subset E^3$, the norm in the corresponding fractal continuum $\Phi_D^3 \subset E^3$ can be defined as follows

$$\|A\| = \left[ \sum_i^3 \chi_{ai}^{2\gamma} \right]^{1/2\gamma}, \tag{32}$$

where the fractional dimension $d_f = 3\gamma \le 3$ is assumed to be the minimal fractional dimension of the space $F^\gamma$ in which the fractal domain $\Phi_3^D$ can be embedded [41]. One may expect (see Ref. [39-41]) that in many cases, but not always, $d_f = d_\ell$ [84]. Notice that, if a path-connected fractal with $2 \le d_\ell < 3$ obeys the Mandelbrot rule of thumb (5), the norm (32) in the corresponding fractal continuum with $\zeta_i \equiv 1$ and $2/3 \leq \gamma = d_\ell/3 < 1$ coincides with the norm (31) in the embedding fractional dimensional space $F^\gamma$. On the other hand, if a fractal domain can be treated as the Cartesian product (6), the norm (32) converts into the norm defined by Eq. (20) for the fractal continuum (16) with $\zeta_i = \alpha_i = d_{\min}^{(i)}$, whereas the norm in the fractal continuum model of the fractal domain (9) (see Fig. 4c) is defined by Eq. (32), but with $d_f = D < d_\ell = 3$, while $\zeta_i$ are given by Eq. (13).

Using the norm (32) the distance between two points $A, B \in \Phi_3^D \subset E^3$ can be defined as

$$\Delta(A,B) = \left[\sum_i^3 \Delta_i^{2\gamma}\right]^{1/2\gamma}, \tag{33}$$

where

$$\Delta_i = |\chi_{ai} - \chi_{bi}| = \xi_0^{1-\zeta_i} \left| sign(x_{ai}) |x_{ai}|^{\zeta_i} - sign(x_{bi}) |x_{bi}|^{\zeta_i} \right|, \tag{34}$$

while $\zeta_i$ are defined by Eq. (13). It is a straightforward matter to verify that the distance defined by Eqs. (33) and (34) satisfies all conventional criteria required of metrics if $1.5 \leq d_f \leq 3$ and $0 < \zeta_i \leq 1$. That is: a) $\Delta(A,B) \geq 0$, b) $\Delta(A,B) = \Delta(B,A)$, c) $\Delta(A,A) = 0$, d) if $\Delta(A,B) = 0$ than $A = B$, e) the triangle inequality $\Delta(A,B) + \Delta(A,C) \geq \Delta(B,C)$. Accordingly, the problems for heterogeneous materials with fractal domains $\Phi_3^D \subset E^3$ with the intrinsic fractal dimension $2 \leq d_\ell \leq 3$ can be mapped into the corresponding problems for the fractal continuum $\Phi_D^3 \subset E^3$ endowed with the norm (32), the metric defined by Eqs. (33) and (34), and the fractal measure defined by Eq. (25) together with relations (27), as it is shown in Fig. 6. Specifically, for the Menger sponge $d_f = d_\ell = D$; for the fractal shown in Fig. 4c $d_f = D < d_\ell = 3$ and $\zeta_i$ are given by Eq. (13), whereas in the case of fractal continuum (16), the metric defined by Eqs. (33) and (34) converts into the metric defined by Eqs. (19) and (21).

### 3.2. Local derivative and Laplacian in fractal continuum

Making use of the fractal metric (34), the local partial derivative in $\Phi_D^3 \subset E^3$ can be defined in a standard manner as follows:

$$\nabla_i^H f = \lim_{\chi_i \to \chi'_i} \frac{f(\chi'_i) - f(\chi_i)}{\chi'_i - \chi_i} = \lim_{x_i \to x'_i} \frac{f(x'_i) - f(x_i)}{\Delta(x'_i, x_i)} = \frac{1}{c_1^{(i)}} \left( \frac{\partial}{\partial x_i} f \right)_{A_2^{(i)} = const}, \quad (35)$$

where the fractal coordinate $\chi_i$ is defined by Eq. (28), $\partial / \partial x_i$ denotes the conventional partial derivative, and $c_1^{(i)}(x_i)$ is given by Eq. (29) which converts into Eq. (24) in the case of fractal continuum (16). In [38] we have recognized that definition (35) formally coincides with the heuristic definition of the *Hausdorff derivative* suggested by Chen [85] and so we have adopted this name. It is also noteworthy to note that the Hausdorff derivative is intimately linked with the scale dependent density of admissible states in the fractal continuum [86]. Accordingly, the Hausdorff derivative resembles, but differs from fractional differential operators used in Refs. [35,36].

The Hausdorff *del* operator in the fractal continuum can be defined in a straightforward manner as

$$\vec{\nabla}^H = \vec{e}_1 \nabla_1^\zeta + \vec{e}_2 \nabla_2^\zeta + \vec{e}_3 \nabla_3^\zeta \,, \qquad (36)$$

where $\vec{e}_i \in E^3$ are base vectors [39], while the norm of a vector is defined by Eq. (32) which coincides with norm defined by Eq. (20) in the special case of fractal continuum (16). The vector fractional differential calculus based on the Hausdorff del operator (36) was developed in Refs. [39,40]. Specifically, it was shown that the Green–Gauss divergence theorem for the fractal continuum reads as

$$\int_A \vec{f} \cdot \vec{n} \, d\mathrm{A}_\partial = \int_W div_H \, \vec{f} \, dV_D \,, \qquad (37)$$

where $\vec{f} = f_k \vec{e}_k$ is any vector field accompanied by the flow through the area $\mathrm{A}_\partial$, $\vec{n} = n_k \vec{e}_k$ is a vector of normal, while

$$div_H f = \vec{\nabla}^H \cdot \vec{f} \ \text{ and } \ rot_H \vec{f} = \vec{\nabla}^H \times \vec{f} \qquad (38)$$

are the Hausdorff divergence and curl operators in $_C \Phi_D^3 \subset E^3$, respectively.

The construction of the Laplacian on fractal domains and in fractional spaces was widely discussed in literature [31,64,79-82,87,88]. Although, generally, the fractional Laplacian cannot be presented as the sum of all the unmixed second partial derivatives [79-82], in the case of fractal continuum (16) the Hausdorff Laplacian can defined in a straightforward way as $\Delta_H = \vec{\nabla}^H \cdot \vec{\nabla}^H$ [39]. On the other hand, Stillinger [79] has phenomenologically introduced the Laplacian $\Delta^F$ in the fractional dimensional space $F^\gamma$. Latter, Palmer and Stavrinou [81] have generalized the Stillinger's Laplacian into the Cartesian coordinates. Further, this Laplacian was employed to solve some physical problems for low dimensional systems and in fractional dimensional spaces (see, for example, Refs. [32,80] and references therein).

Following to Refs. [79,81], the generalized fractional Laplacian in $\Phi_D^3 \subset F^\gamma \subset E^3$ can be phenomenologically defined in the following form

$$\Delta_H^F f = \sum_i^3 \left(c_1^{(i)}\right)^{-2} \left[\left(\frac{\partial^2 f}{\partial x_i^2}\right) + \frac{\gamma - \zeta_i}{x_i}\left(\frac{\partial f}{\partial x_i}\right)\right] \tag{39}$$

accounting for the fractal topology of the fractal domain with $d_\ell = 3\gamma$, as well as its fractional metric (34) [40,41]. Notice that when $\zeta_i = D - D_S^{(i)} \equiv 1$, Eq. (39) coincides with the generalized Laplacian $\Delta^F$ in an isotropic fractional space $F^\gamma$ (see Ref. [81]), whereas if $d_f = d_\ell = 3$ but $\zeta_i \neq 1$ Eq. (39) converts into the Hausdorff Laplacian $\Delta_H = \vec{\nabla}^H \cdot \vec{\nabla}^H$ for the fractal continuum (16) which, in the limit of $\gamma = \zeta_i = D - D_S^{(i)} \equiv 1$, converts into the conventional Laplacian in the Euclidean space, even when $D < 3$.

Furthermore, to deal with diffusion processes and strain (stress) relaxation in fractal materials, one may need to use the *Hausdorff time derivative* introduced in Ref [84], while the order of Hausdorff time derivative is determined by the spectral dimension of the fractal domain under study [39-41].

## IV. Mechanics of fractal continuum

The mechanics of deformable materials cannot be deduced from the laws of mechanics of material points and rigid bodies. Additional assumptions must be introduced to define the notions of internal and external stresses, and the equilibrium equation should be defined. The geometric framework in which both the classical and the fractal continuum mechanics are worked is the three-dimensional Euclidean space $E^3$. Both approximate physical realities only when the properties of fractal materials are studied in a smoothed picture. Hence, the deformations and stresses which can be considered in the fractal continuum are only those produced during the application of external forces. Accordingly, to develop the fractal continuum mechanics we need first to develop the kinematics of deformations, define the notion of stresses, and then establish the balance (conservation) and constitutive laws for fractal continua.

In this context, it is pertinent to note that some kinds of deformations of a material manifold may lead to change its metric [89,90]. Evolving metrics have been extensively studied in mathematics [91]. Furthermore, in [92] was developed a geometric theory of thermo-elasticity in which thermal strains are buried in a temperature-dependent

Riemannian material manifold, such that a change of temperature leads to a rescaling of the material metric with a clear physical meaning. A geometric theory of growth mechanics with evolving metric was suggested in [93]. Although, the evolution of fractional metric can be also considered within a fractal continuum framework, below we explicitly assume that deformations considered in this work do no lead to a change of fractal metric in the deformed fractal continuum.

### 4.1. Kinematics of fractal continuum deformation

Once again, let us first consider the special case of fractal domain (6). In the corresponding fractal continuum (16) there is a direct correspondence between the fractal and Cartesian coordinates. Accordingly, let us at time $t = 0$ the fractal domain $_C\Phi_3^D \subset E^3$ occupies region $W_0(\bar{X} \in E^3) \subset E^3$ and, at time $t > 0$, occupies a region $W_t(\bar{x} \in E^3) \subset E^3$, such that the corresponding fractal continuum (16) occupies region $W_0(\bar{X} \in {}_C\Phi_D^3) \subset E^3$ at $t = 0$ and region $W_t(\chi \in {}_C\Phi_D^3) \subset E^3$ at $t > 0$ (see Fig. 7). In both cases, the initial and current configurations are supposed to be bounded, open, and connected. Therefore, the motion of fractal continuum can be determined either by the current position $\chi \in {}_C\Phi_D^3$ of the material point in $_C\Phi_D^3 \subset E^3$ as a function of the initial (reference) position $\bar{X} \in {}_C\Phi_D^3$ and time $t$, or by the current position $\bar{x} \in E^3$ of the same material point as a function of the reference position $\bar{X} \in E^3$ and $t$. Specifically, the displacement vector $\upsilon = (\upsilon_1, \upsilon_2, \upsilon_3) \in {}_3\Phi_D^3 \subset E^3$ describing the displacement field in the reference configuration is equal to

$$\upsilon_i = \chi_i(\mathrm{X}_i, t) - \mathrm{X}_i = \xi_0^{1-\varsigma_i}\left[(x_i + u_i)_i^{\varsigma_i} - X_i^{\varsigma_i}\right], \tag{40}$$

where $u_i = (x_i - X_i) \in E^3$ are components of the displacement vector in the embedding Euclidean space (see Fig. 7), $\varsigma_i = \alpha_i$, and the length of displacement vector $\vec{\upsilon} \in \Phi_D^3$ is defined by the norm (20). For simplicity, in Eq. (40) and everewhere below in this paper we use the notation

$$\chi_i = sign(x_i)\xi_0^{1-\varsigma_i}|x_i|^{\varsigma_i} \equiv \xi_0^{1-\varsigma_i}x_i^{\varsigma_i}, \tag{41}$$

where $-\infty < \chi_i < \infty$, while $-\infty < x_i < \infty$.

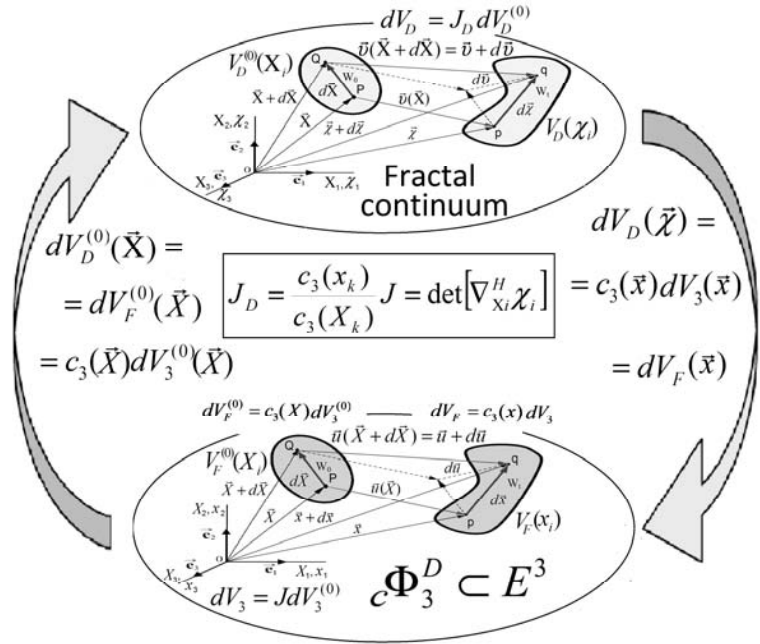

**Figure 7.** Mapping of deformable fractal medium into the fractal continuum and the corresponding transformations from the original (reference) to deformed (current) configurations [see Eq. (43)].

It is a straightforward matter to understand that the Jacobian matrix in the fractal continuum should be defined as the deformation gradient $\bar{F}_D = \left[\partial\chi_i/\partial X_j\right]$ in $_C\Phi_D^3 \subset E^3$ and so, the Jacobian of transformation in the fractal continuum takes the following form

$$J_D = \det\left[\nabla_{Xi}^H\chi_i\right] = \frac{c_3(x_k)}{c_3(X_k)}\det\left[\nabla_{Xi}x_i\right] = \frac{c_3(x_k)}{c_3(X_k)}J,\qquad(42)$$

where $J = \det\left[\nabla_{Xi}x_j\right]$ denotes the conventional Jacobian of transformation in $E^3$, such that $dV_3 = JdV_3^{(0)}$ [41]. Consequently, the infinitesimal volume element of $W_t \subset {}_C\Phi_D^3$ transforms as

$$dV_D = c_3(\bar{x})dV_3 = c_3(\bar{x})JdV_3^{(0)} = c_3(\bar{x})c_3^{-1}(\bar{X})JdV_D^{(0)} = J_D dV_D^{(0)},\qquad(43)$$

as it is illustrated in Fig. 7. So, deformations conserve the volume of region $W_t \subset {}_C\Phi_D^3$ if and only if $J_D = 1$. Furthermore, for every $t > 0$, function $\bar{\chi}(\bar{X}, t)$ is a smooth one-to-one map of every material point of $W_0(\bar{X}) \subset {}_C\Phi_D^3$ onto $W_0(\bar{\chi}) \subset {}_C\Phi_D^3$, such that there exists a unique inverse of (40), at least locally, if and only if $J_D$ is not identically zero, that is $0 < J_D < \infty$ [41].

Consequently, the Lagrangian (Green) strains in the fractal continuum $_C\Phi_D^3 \subset E^3$ are defined as $E_{ij} = 0.5\left(\nabla_{Xi}^H\upsilon_i + \nabla_{Xj}^H\upsilon_j + \nabla_{Xj}^H\upsilon_k\nabla_{Xi}^H\upsilon_k\right)$, whereas the Eulerian (Almansi) strain tensor takes the form $e_{ij} = 0.5\left(\nabla^H_j \upsilon_i + \nabla^H_i \upsilon_j - \nabla^H_j \upsilon_k \nabla^H_i \upsilon_k\right)$ [41]. In the limit of infinitesimally small deformations (see), both tensors are converted into the infinitesimal strain tensor [41]:

$$\varepsilon_{ij} = \frac{1}{2}\left(\nabla^H_j \upsilon_i + \nabla^H_i \upsilon_j\right), \qquad (44)$$

where $\upsilon_i$ are defined by Eq. (34), as it is shown in Fig. 8.

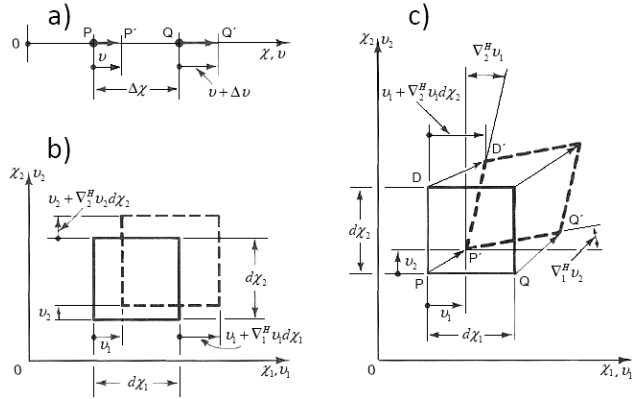

**Figure 8.** Geometry and metric of: (a,b) normal and (c) shear strains in the fractal continuum.

In the case of fractal domains $\Phi^D_3 \subset E^3$ having the intrinsic fractal dimension $2 \le d_\ell < 3$, only two mutually orthogonal fractal coordinates can be defined. Accordingly, in the corresponding fractal continuum $\Phi^3_D \subset E^3$ we can choose any pair ($\chi_i = \chi$, $A^{(i)}_\partial = A_\partial$) as it is defined by Eqs. (27)-(29). For example, for the Menger sponge (see Fig. 6), from Eq.

(30) follows that in the initial configuration of $\Phi_D^3 \subset E^3$ we have $X = \xi_0^{1-\zeta} X^{\zeta}$ and $A_\partial^{(0)} = \xi^{2-D_S} |X_2 X_3|^{D_S/2}$, whereas in the current configuration

$$\chi = \xi_0^{1-\zeta} (X - u_1)^{\zeta} \text{ and } A_\partial = \xi^{2-D_S} |(X_2 + u_2)(X_3 + u_3)|^{D_S/2}, \tag{45}$$

where $u_i = (x_i - X_i)$ are components of the displacement vector in $E^3$. Accordingly, to construct the Jacobian matrix for coordinate changes, in addition to coordinate pair ($X$, $A_\partial^{(0)}$) we need to introduce a fictitious auxiliary coordinate $Z^{(0)}$ without any physical meaning (see, for example, Ref. [94,95]), which is not changed during the fractal continuum deformation (that is $Z(t) = Z^{(0)}$). Consequently, the Jacobian matrix takes the following form

$$J_D^{\chi A} = \begin{bmatrix} \dfrac{\partial \chi}{\partial X} & \dfrac{\partial \chi}{\partial A_\partial^{(0)}} & 0 \\ \dfrac{\partial A_\partial}{\partial X} & \dfrac{\partial A_\partial}{\partial A_\partial^{(0)}} & 0 \\ 0 & 0 & 1 \end{bmatrix} \tag{46}$$

and so once again, the Jacobian of transformation in $\Phi_D^3 \subset E^3$ can be presented as

$$J_D = \det \left[ J_D^{ij} \right] = \frac{c_1(x_i) c_2^{(i)}(x_j)}{c_1(X_i) c_2^{(i)}(X_j)} J = \frac{c_3(x_k)}{c_3(X_k)} J, \tag{47}$$

where the equality (26) is employed, while $J$ is the conventional Jacobian of transformation in $E^3$, such that

$$dx_i = (\partial x_i / \partial X_t) dX_i , \ dA_2^{(i)} = J\left|\bar{F}^{-T} n_i^{(0)}\right| dS_2^{(0)} , \text{ and } dV_3 = dX_i dA_2^{(0)} = JdV_3^{(0)} ,$$

where $\bar{F} = \left[\partial x_i / \partial X_i\right]$ is the deformation gradient in $E^3$, $\bar{F}^{-T}$ denotes the inverse transpose of $\bar{F}$, and $\bar{n}_i^{(0)}$ is the unit normal vector in the initial configuration, whereas in the current configuration $\bar{n}_i = \bar{F}^{-T}\bar{n}_i^{(0)} \big/ \left|\bar{F}^{-T}\bar{n}_i^{(0)}\right|$ [96]. Therefore, the infinitesimal volume element of $W_t(\chi, \mathrm{A}) \subset \Phi_D^3$ transforms as follows

$$
\begin{aligned}
dV_D = d\chi d\mathrm{A}_\partial &= c_1 c_2 dx_1 dA_2 = c_3(\bar{x}) JdV_3^{(0)} = c_3(\bar{x}) JdX dA_2^{(0)} \\
&= c_3(\bar{x}) c_1^{-1} c_2^{-1} JdX d\mathrm{A}^{(0)} = c_3(\bar{x}) c_3^{-1}(\bar{X}) JdV_D^{(0)} = J_D dV_D^{(0)}
\end{aligned}
\tag{48}
$$

and so deformations conserve the volume of region $W_t \subset \Phi_D^3$ if and only if $J_D = 1$. In this regard, it should be pointed out that, although in any fractal continuum three components ($\upsilon_i$) of displacement vector $\bar{\upsilon}$ are defined by Eq. (40), the length of displacement vector is generally controlled by the norm (32), which converts into the norm (20) only in the case (16). Nonetheless, the infinitesimal strain tensor in the fractal continuum with $2 < d_\ell < 3$ has the form of Eq. (44).

### 4.2. Euler's identity, material derivative, Reynolds transport theorem and continuity equation for fractal continuum

Using the conventional rule for determinant differentiating, from Eq. (47) immediately follows the generalized Euler's identity for fractal continuum $\Phi_D^3 \subset E^3$ in the following form

$$\left(\frac{dJ_D}{dt}\right)_D = J_D \nabla_i^H v_i,\tag{49}$$

where $v_i = \partial v_i / \partial t$ are components of the velocity vector in the fractal continuum and $(d/dt)_D$ denotes the fractal material (Lagrangian) derivative, which in $\Phi_D^3 \subset E^3$ has the form

$$\left(\frac{d\psi}{dt}\right)_D = \frac{\partial}{\partial t}\psi + v_i \nabla_i^H \psi,\tag{50}$$

where $\psi(x_i, t)$ is any extensive quantity accompanied by a moving region $W_t \subset \Phi_D^3$ and the usual summation convention over repeated indices is assumed [41]. Consequently, the Reynolds' transport theorem for fractal continuum reads as

$$\left(\frac{d}{dt}\right)_D \int_{W_t} \psi dV_D = \int_{W_t} \left(\partial \psi / \partial t + \nabla_k^H (\psi v_k)\right) dV_D = \int_{W_t} \left(\partial \psi / \partial t\right) dV_D + \int_{\partial W} \psi v_k n_k d\mathrm{A}_D^{(k)},\tag{51}$$

where the first term on the right hand side is the time rate of change of $\psi$ within the control volume of $W_t$ and the second term represents the net flow of $\psi$ across the control surface $\partial \Phi \in \Phi_D^3$ of region $W_t \in \Phi_D^3 \subset E^3$ (see <span style="color:red">Fig. 9</span>). Furthermore, using the fractal material derivative (50), the equation of mass conservation can be presented as follows:

$$\partial \rho_c / \partial t = -div_H (\rho_c \vec{v}),\qquad(52)$$

where $\rho_c$ is the overall mass density of the fractal continuum [41].

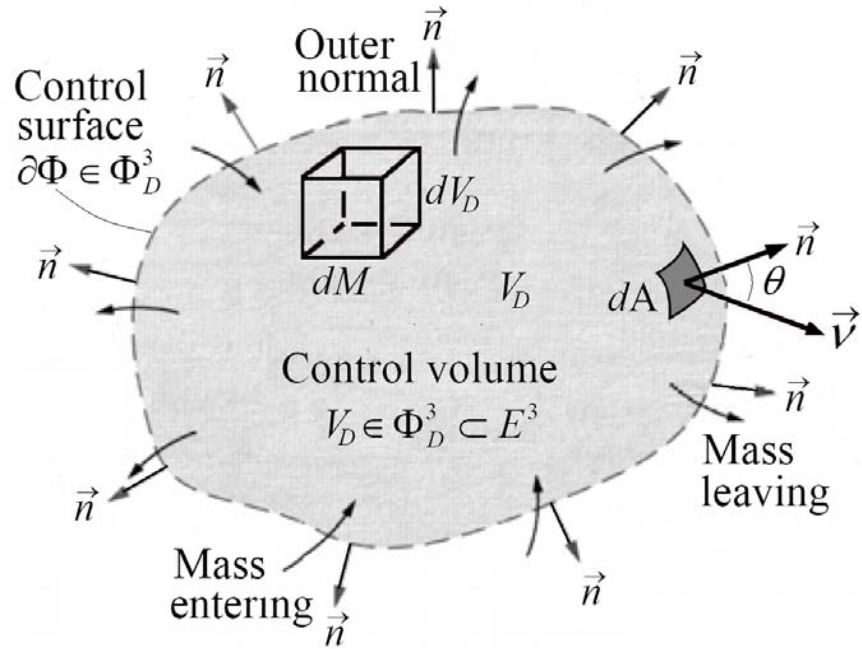

**Figure 9.** Illustration of the Reynolds transport theorem for the fractal continuum. The integral over the control surface $\partial \Phi$ gives the net amount of the property $\psi$ following out of the control volume (into the control volume, if it is negative), per unit time [see Eq. (51)].

## 4.3. Forces and stresses in fractal continuum

Following the paradigm of classical continuum mechanics [97], the forces that act on the fractal continuum or its part can be divided into two categories: those that act by contact with the surface ($\vec{F}_s$), called surface tractions, and those that act at a distance ($\vec{F}_b$), termed as the volume or body forces. If $dF_i$ is a contact force acting on the deformed area $dA_\partial^{(i)} = \vec{n}_i\, dA_\partial$, where $\vec{n}$ is the unit outer normal to the element of area $dA_\partial \in \Phi_D^3$, then the stress (traction) vector can be defined as

$$\vec{t}_{\vec{n}}(\vec{x},t) = \lim_{\Delta A_\partial \to 0} \frac{\Delta \vec{F}(\vec{x},t)}{\Delta A_\partial} = \lim_{\Delta A_2 \to 0} c_2^{-1} \frac{\Delta \vec{F}}{\Delta A_2}. \tag{53}$$

where $c_2 = \left(dA_\partial / dA_2\right)$ is the density of admissible states in the area $A_\partial \in \Phi_D^3$ (e.g., for the Menger sponge $c_2$ is given by the second relationship in Eq. (30)]. Accordingly, assuming that this limit exists, one can define the normal and shear stresses in the usual way [41]. Generally, stresses are not uniformly distributed over a fractal continuum, and may vary with time. Therefore the stress tensor must be defined for each point and each moment, by considering an infinitesimal particle of the medium surrounding that point, and taking the average stresses in that particle as being the stresses at the point (see Fig. 10). Accordingly, the total force acting on the fractal continuum $\Phi_D^3 \subset E^3$ can be presented in the general form as

$$\vec{f} = \int_\Phi \vec{f}_b\, dV_D + \int_{\partial\Phi} \vec{t}_{\vec{n}}\, dA_\partial, \tag{54}$$

whereas the deformations of fractal continua should satisfy the laws of momentum conservation allied with the Newton's first and second laws [44].

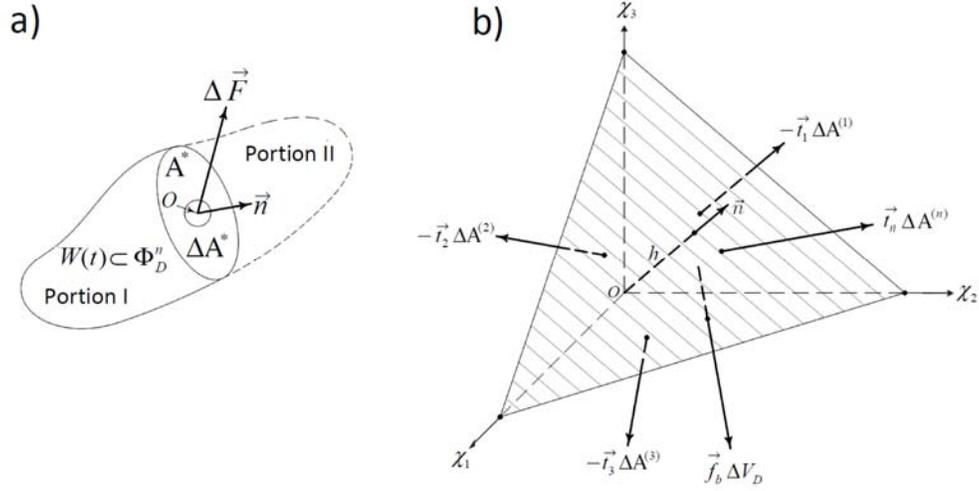

**Figure 10**. (a) Surface force on surface element $\Delta A^*$ and (b) equilibrium of an infinitesimal tetrahedron in $\Phi_D^3 \subset E^3$.

The principle of linear momentum balance states that the time rate of change of the linear momentum is equal to the resultant force acting on the body. Hence, the principle of linear momentum conservation in fractal continuum implies that

$$\iint\limits_{W_t} \left( \nabla_j^H \cdot \sigma_{ij} + \vec{f}_b - \rho_c \partial^2 \vec{\upsilon}/\partial t^2 \right) dV_D = 0,$$ (55)

where $\partial^2 \vec{\upsilon}/\partial t^2$ is the acceleration field. Using the continuity equations (52), the law of linear momentum conservation in fractal continuum can be presented in the local form as

$$\rho_c \partial^2 \upsilon_i / \partial t^2 = f_b^{(i)} + \nabla_j^H \cdot \sigma_{ij}, \qquad (56)$$

where $\sigma_{ij}$ is the Cauchy stress tensor (see Fig. 11). Notice that Eq. (56) converts into the conventional equation for the density of linear momentum balance if the fractal dimension of any intersection is equal to $D_S^{(i)} = D - 1$ [41].

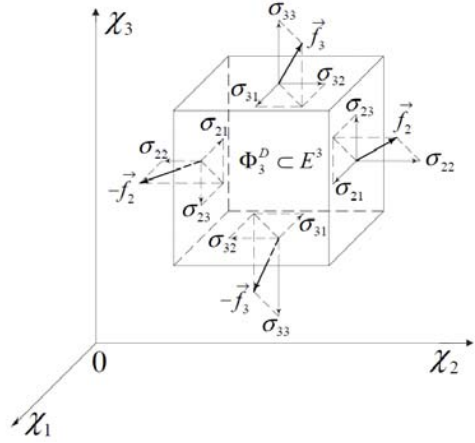

**Figure 11.** The components of stress tensor in the fractal continuum $\Phi_D^3 \subset E^3$.

The principle of angular momentum balance asserts that the time rate of change of the moment of momentum of a body with respect to a given point is equal to the moment of the surface and body forces with respect to that point:

$$(d/dt)_D \int\limits_{W_t} \rho_c e_{ijk} \chi_i v_j dV_D = \int\limits_W e_{ijk} \varsigma_0^{1-\zeta_i} x_i^{\zeta_i} f_j dV_D + \int\limits_{\partial W} e_{ijk} \varsigma_0^{1-\zeta_i} x_i^{\zeta_i} \sigma_{mj} n_m dA_D, \qquad (57)$$

where $e_{ijk}$ is the permutation tensor. Consequently, on account of the Green-Gauss (37) and Reynolds's transport (51) theorems, the law of angular momentum conservation in the fractal continuum can be presented in the local form $e_{ijk}\sigma_{ij}=0$ from which immediately follows that in the absence of any internal angular momentum, body couples, and couple stresses the Cauchy stress tensor in the fractal continuum is symmetric, that is

$$\sigma_{ij}=\sigma_{ji}\,,\qquad\qquad(58)$$

regardless of whether or not it is in equilibrium [44].

### 4.4. Constitutive laws for fractal continuum

The constitutive laws of solid mechanics cannot be deduced from the general laws of continuum mechanics, and so they are defined from physical experiments (for example, Hooke's law of elasticity, micropolar elasticity law, visco-elastic law, strain-hardening plasticity law, etc.). In this way, there are a number of rules that must be fulfilled to establish a constitutive equation that is admissible from the rational and physical standpoints [98]. Specifically, constitutive equations should be invariant under any change of reference frame. Furthermore, the current rheological and thermodynamic state of the material should be completely determined by the history of the thermo-kinetic process experienced by the material. In the case of incompressible materials the stress state is determined to within the hydrostatic pressure, which depends on the boundary conditions and the problem geometry. As well, the stress tensor at a given point does not depend on movements occurring at finite distance from this point.

In the case of solid materials with (pre-)fractal (micro-)structure the constitutive equations are dependent of the mechanical properties of matter, as well as on the fractal features of the (micro-)structure [12-14]. Accordingly, the constitutive laws for deformable fractal continua can be defined by the mapping of classical constitutive relations into the fractal continuum framework. Specifically, the constitutive law for linear elastic isotropic ($\zeta_i \equiv \zeta$) fractal continuum takes the following form

$$\sigma_{ij} = \mu\left(\nabla_j^H \upsilon_i + \nabla_i^H \upsilon_j\right) + \lambda \nabla_k^H \upsilon_k \delta_{ij} = 2\mu\varepsilon_{ij} + \lambda\varepsilon_{kk}\delta_{ij}, \qquad (59)$$

where the deformation tensor $\varepsilon_{ij}$ is defined by Eq. (44), while $\lambda$ and $\mu$ are the effective Lame coefficients of the fractal continuum [41]. Generalizations for anisotropic linearly elastic and elastoplastic fractal continua are also straightforward.

## V. Mapping of mechanical problems for fractal materials into problems for fractal continuum

### 5.1. Stresses and strains in the fractal bars

Mechanics of fractal bars has attached an increasing interest in the physics and material sciences [16,99,100]. Here, let us first consider a slender bar with a fractal microstructure of weight $M = \rho_{bar} L A_2$ and cross-sectional area $A_2 = B^2$ that is suspended from a ceiling (see Fig. 12a), where $\rho_{bar}$ is the overall density of bar. To determine the normal force caused by the weight of the bar, the problem can me mapped into the problem for the fractal continuum bar of the same mass $M = \rho_C \xi_0^{1-\zeta} L^\zeta A$, cross-sectional area $A = c_2(A)A_2$, and the mass density equal to

$$\rho_C = \rho_{bar}\left(\frac{B}{\xi_0}\right)^{2-D_s}\left(\frac{L}{\xi_0}\right)^{1-\zeta} = \rho_0 > \rho_{bar} = \rho_0\left(1-\phi\right), \tag{60}$$

where $\rho_0$ is the density of material from which the bar is made and $\phi$ is its overall porosity (see Fig. 12b). If we cut the fractal continuum bar at an arbitrary position $\chi$, the normal force $F_n(\chi)$ is equal to the weight of the portion of the bar below the imaginary cut. That is $F_n = gM(1-\chi/\Lambda)$ and so, the normal stress is equal to

$$\sigma_n(\chi) = \frac{F_n}{A} = \frac{Mg}{A}\left(1-\frac{\chi}{\Lambda}\right), \tag{61}$$

where $g$ is the gravitational acceleration constant and $\Lambda = \xi_0(L/\xi_0)^\zeta$. Accordingly, in the fractal coordinates the stress linearly decreases from $\sigma_n(\chi = 0) = M/A_\partial$ to $\sigma_n(\chi = \Lambda) = 0$ at the free end. The strain distribution in the elastic fractal continuum model is $\varepsilon_n = \sigma_n/E$, where $E$ is the Young modulus of fractal continuum equal to the Young modulus of the fractal bar. Consequently, the displacement in the elastic fractal continuum bar behaves as

$$v_n(\chi) = \frac{gM\Lambda}{E\mathrm{A}} \left( \frac{\chi}{\Lambda} - \frac{\chi^2}{2\Lambda^2} \right) \tag{62}$$

and so, the elongation of the fractal continuum bar due to its own weight is

$$\Delta\Lambda = \int_0^\Lambda \frac{F_n(\chi)}{E\mathrm{A}} d\chi = \frac{gM}{E\mathrm{A}} \int_0^\Lambda \left( 1 - \frac{\chi}{\Lambda} \right) d\chi = \frac{1}{2}\frac{gM}{E}\frac{\Lambda}{\mathrm{A}}. \tag{63}$$

The mapping $\Phi_D^3 \to \Phi_3^D \subset E^3$ implies that the elongation of the fractal bar in $E^3$ due to its own weight

$$\Delta L = \frac{gM}{2EB^2} \left( \frac{\xi_0}{L} \right)^{1-\zeta} \left( \frac{B}{\xi_0} \right)^{2-D_s} \tag{64}$$

can be either larger, or smaller than the elongation of the homogeneous bar of the same mass ($M$), Young modulus ($E$), and overall sizes ($L$, $B$). Specifically, if

$$\left( \frac{B}{\xi_0} \right)^{2-D_s} < \left( \frac{L}{\xi_0} \right)^{1-\zeta} \tag{65}$$

the elongation of the fractal bar is less than the elongation of the homogeneous one and vice versa. Furthermore, from Eq. (61) it follows that the overall stress distribution in the fractal bar

$$\sigma_n(x) = \frac{gM}{\xi_0^{2-D_S}B^{D_S}}\left(1 - \frac{x^\zeta}{L^\zeta}\right) \qquad (66)$$

is non-linear. Consequently, the overall longitudinal strain in the fractal bar (Fig. 12a) behaves as

$$\varepsilon_n = \frac{g\rho_0 L}{E}\left(\frac{\xi_0}{L}\right)^{1-\zeta}\left(1 - \frac{x^\zeta}{L^\zeta}\right), \qquad (67)$$

such that at the free end $\varepsilon_n(x = L) = \varepsilon_n(\chi = \xi_0^{1-\zeta}L^\zeta) = 0$ (Fig. 12c), whereas at the upper end ($x = 0$):

$$\varepsilon_n = \frac{gM}{EA} = \frac{g\rho_0\xi_0}{E}\left(\frac{L}{\xi_0}\right)^\zeta = \frac{g\rho_{bar}L}{E}\left(\frac{B}{\xi_0}\right)^{2-D_S}. \qquad (68)$$

Hence, if the fractal and homogeneous bars have the same overall sizes ($L$ and $B$), Young modulus ($E$), and the overall density ($\rho_{bar} = M/LB^2$), the stress (66) and strain (68) at the upper end of fractal bar are larger than in the homogeneous one, even when the overall elongation of the fractal bar (64) is less than the overall elongation of the homogeneous bar. This is easy to understand taking into account that the force acting at the upper end is the same $F_n = gM$, whereas the effective area of fractal bar intersection (with $D_S < 2$) is less then the area of the homogeneous bar intersection. However, in the special case of $D_S = 2$, when $A = A = B^2$, the strains at the upper ends of fractal and homogeneous bars are equal, even when $D < 3$ and $\zeta = D - 2 < 1$, whereas the overall elongation of the fractal bar with $D_S = 2$ is always less than the elongation of the homogeneous bar.

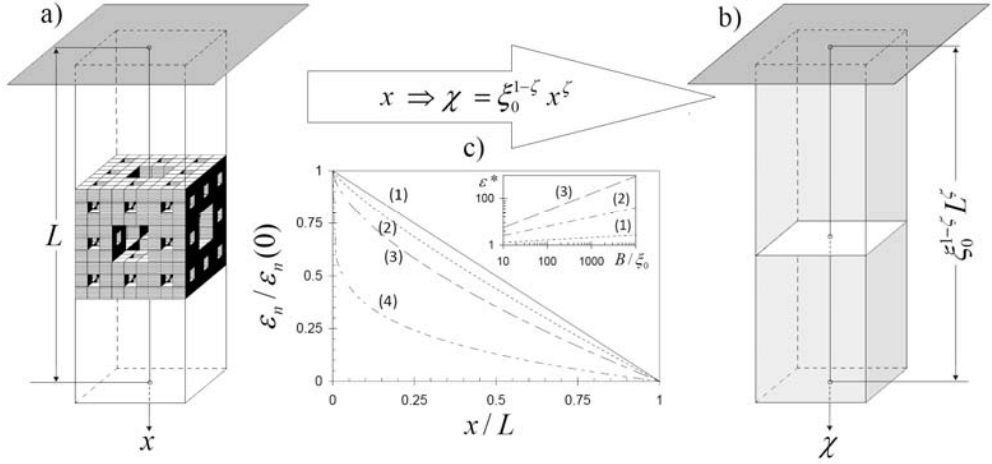

**Figure 12**. Mapping of the fractal bar (a) into the bar made of fractal continuum (b) and graphs of $\varepsilon_n(x)/\varepsilon_n(0)$ versus $x/L$ (c) for: homogeneous bar ($\zeta = 1$) (1); and fractal bars with: $\zeta = \ln 2.5 / \ln 3 = 0.834$ (2); $\zeta = \ln 2 / \ln 3 = 0.63$ (3); and $\zeta = 0.2$ (4). Inset shows the graphs of $\varepsilon^* = \varepsilon_n(0) E / g \rho_{bar} L$ versus $B/\xi_0$ for fractal bars with $D_S = \ln 8 / \ln 3 = 1.89$ (1); $D_S = 1.6$ (2); $D_S = \ln 4 / \ln 3 = 1.26$ (3).

Now, let us determine the strains in the elastic fractal bar subjected to tensional force $F_n \gg gM$ (see Fig. 13a). The problem can be mapped into the problem for the fractal continuum bar with the boundary condition $\upsilon_n(0) = 0$ (see Fig. 13b). Accordingly, the overall stress (59) in the fractal bar

$$\sigma_n = E \nabla_1^H \upsilon = E \frac{F_n}{A} \qquad (69)$$

is constant. Consequently, the overall strain

$$\varepsilon_n = \nabla_1^H \upsilon_n = F_n / A \tag{70}$$

is also constant along the fractal bar, as this is in the homogeneous bar. Furthermore, from Eq. (70) follows that the fractal displacement is $\upsilon_n = (F_n / EA)\chi$ and so the apparent displacement distribution in the Euclidean coordinates is non-linear. Namely,

$$u_n = \frac{F_n x}{EA} \left(\frac{\xi_0}{B}\right)^{2-D_S} \left(\frac{\xi_0}{x}\right)^{1-\zeta} \tag{71}$$

and so the displacement at the free end of the fractal bar

$$\Delta L = u_n(x = L) = \frac{F_n L}{EB^2} \left(\frac{B}{\xi_0}\right)^{2-D_S} \left(\frac{\xi_0}{L}\right)^{1-\zeta} \tag{72}$$

can be either less or larger the elongation of the homogeneous bar (see Fig. 13c). Specifically, if the relation (65) holds, the effective rigidity of the fractal bar is larger than the rigidity of the homogeneous bar of the same overall size, mass, and Young modulus.

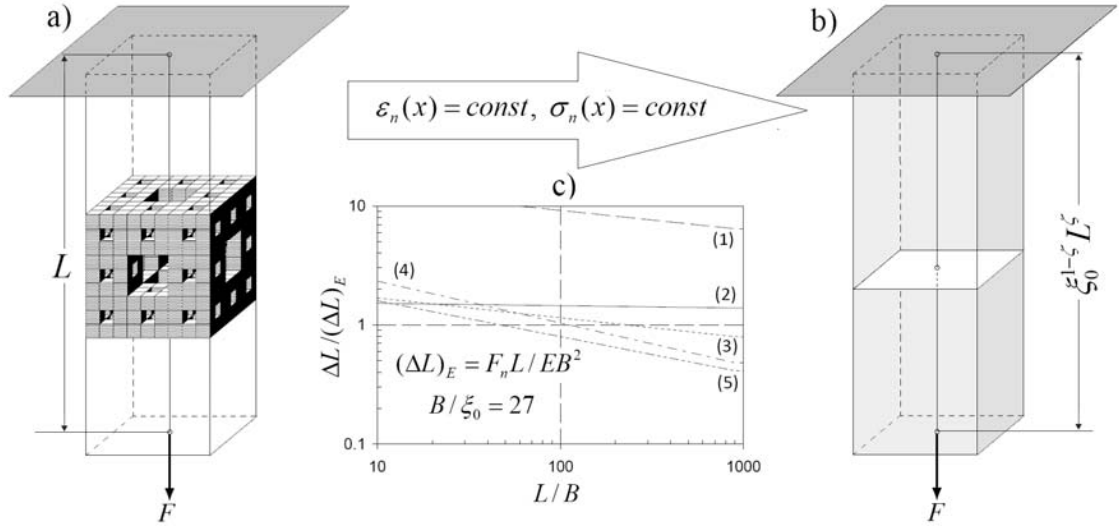

**Figure 13**. Mapping of the problem for fractal bar subjected to the tensile force $F_n \gg Mg$ (a) into the corresponding problem for fractal continuum (b) and graphs of normalized displacement at the free end $\Delta L/(\Delta L)_E$ versus the ratio $L/B$ (c) for fractal bars ($B/\xi_0 = 27$) with: $D = 2.1$, $D_S = \ln 4/\ln 3$, and $\zeta = D - D_S = 0.838$ (1); $D = 2.86$, $D_S = 1.88$, and $\zeta = 0.98$ (2); Menger sponge ($D = \ln 20/\ln 3 = 2.727$, $D_S = \ln 8/\ln 3 = 1.89$, and $\zeta = \ln 2.5/\ln 3 = 0.834$) (3); $D = 2.5$, $D_S = 1.85$, and $\zeta = 0.65$ (4); $D = 2.86$, $D_S = 1.88$, and $\zeta = 0.98$ (2); (3); $D = 2.65$, $D_S = 1.95$, and $\zeta = 0.7$ (4); and $D = 2.65$, $D_S = 1.95$, and $\zeta = 0.7$ (5), whereas the displacement expected for homogeneous (Euclidean) bar is $(\Delta L)_E = F_n L/EB^2$.

In this way, the fractal continuum homogenization method allows us to optimize the bar structure. Although, as far as we know, there are no available experimental data for quantitative comparison with our theoretical results, we noted that experiments reported in [16] are qualitatively consistent with our finding in a sense that the fractal bar can have larger rigidity that the homogeneous one of the same weight. Furthermore, our theoretical predictions can be easily verified in experiments with fractal bars made, for example, with the help of a 3D printer. So, we expect that our findings will stimulate further experimental efforts in this direction.

## 5.2. Elastic waves in fractal continuum

The elastic wave propagation and localization in fractals and fractal materials is a tremendous importance for both fundamental and technological interest [101,102,103,104,105, 106,107,108,109]. Within a fractal continuum approach, the equation of elastic wave propagation in the fractal domain (7) can be obtained from Eq. (56) with $f_b^{(i)} = 0$ and stresses defined by Eq. (59). This leads to the following wave equation

$$\rho_C \left( \partial^2 \upsilon / \partial t^2 \right) = \mu \Delta_H \upsilon + \mu \left( 1 - 2\nu \right)^{-1} \nabla^H \left( \nabla^H \cdot \upsilon \right), \tag{73}$$

where displacement vector $\upsilon$ is defined by Eq. (40), $\Delta_H = \nabla^H \cdot \nabla^H$ is the Hausdorff Laplacian, and $\nu$ is the Poisson ratio. Notice that Eq. (73) coincides with the conventional wave equation up to the coordinate transformation $x_i \rightarrow \chi_i$, such that $u_i \rightarrow \upsilon_i$. So, the overall solutions of elastic wave propagation problems in the fractal domain (7) can be easily obtained from the solutions of the corresponding problems for a homogeneous Euclidean domain by the inverse coordinate transformations (see Ref. [31b]). Hence, the fractal metric defined by Eqs. (19) and (21) does not cause localization of elastic waves.

A simple heuristic way to account for the fractal topology of fractal domains with $d_\ell < 3$ on the elastic wave propagation is to substitute the operator $\Delta_H = \vec{\nabla}^H \cdot \vec{\nabla}^H$ in Eq. (73) by the Laplacian (39). Accordingly, the wave equation in the fractal continuum takes the form

$$\rho_C \left( \partial^2 \vec{\upsilon} / \partial t^2 \right) = \mu \Delta_H^F \vec{\upsilon} + \mu (1 - 2\nu)^{-1} \vec{\nabla}^H \left( \vec{\nabla}^H \cdot \vec{\upsilon} \right). \qquad (74)$$

It is noteworthy to note that Eq. (74) mathematically resembles the wave equation used in [104] to the study of the elastic wave localization in heterogeneous media. Taking into account the results obtained in Ref. [104], it is a straightforward matter to understand that the fractal topology of medium with the intrinsic fractal dimension $d_\ell = 3\gamma < 3$ causes the localization of elastic waves in it. In this regard, it is pertinent to note that the elastic wave localization in fractal materials was observed as in numerical simulations [104], as well as in experiments reported in Refs. [106,108,110]. Furthermore, it is easy to see that the fractal topology causes the coupling of longitudinal and transverse waves in the fractal continuum, such that in fractal domains with $d_\ell < 3$ the pure transverse waves cannot propagate. We expect that these finding will stimulate further experimental research of elastic wave propagation in fractal media, such that experimental results can be quantitatively compared with the theoretical predictions based on Eq. (74).

### 5.3. Crack tip stress fields in fractal materials

Crack propagation in heterogeneous materials has a great importance in all areas of engineering ranging from nanotechnologies to petroleum industry. Classical fracture mechanics is based on the concept of homogeneous continuum and smooth (Euclidean) geometry of crack path and fracture surfaces [111]. Actually, however, the crack geometry exhibits long-range (self-affine) correlations, even in the absence of long-range correlations in the material microstructure (see Refs. [112,113,114,115,116,117]. The self-affine geometry of a crack leads to the change of the stress distribution in the crack tip vicinity [118,119,120,121,122,123,124,125,126]. On the other hand, the long-range correlations in the material microstructure affect the stress concentration ahead of a straight notch [37,127,128]. Furthermore, the long-range correlations in the material microstructure determine the fractal geometry of admissible cracks [129,130]. So, the crack mechanics in fractal materials should account for the fractal properties of material microstructure, as well as the crack geometry.

The non-differentiability of fractals does not permit to formulate the boundary conditions on the fractal fracture surface in the usual way. Nonetheless, the asymptotics of stress distributions in the vicinity of self-affine crack can be deduced either from the energy balance considerations [116-118,120,122], or by the mapping of a problem with self-affine crack into the problem with a straight crack loaded by unknown traction [119a]. All methods predict that the crack tip stress field in the vicinity of self-affine crack tip ($\sigma \propto r^{-\varsigma}$) is less divergent than ahead of a smooth cut in a homogeneous continuum ($\sigma \propto 1/\sqrt{r}$). Furthermore, the energy balance considerations suggest that the stress concentration at the tip of a straight cut in a fractal material ($\sigma \propto r^{-\varsigma}$) is less divergent ($\varsigma < 0.5$) than in the homogeneous one [125].

The analytic envelope of stress field distribution ahead of a crack in body made of a fractal material can be obtained by the mapping of crack problem for the fractal body into the corresponding crack problem for the fractal continuum. In particular, if the intrinsic fractal dimension of the fractal domain $\Omega_F(\Phi_3^D) \subset E^3$ is $d_\ell = 3 > D > 2$ (e.g., Fig. 4), the distribution of crack tip stresses can be directly obtained from the solution of the corresponding elastic crack problem for the Euclidean continuum by the coordinate change $x_i \rightarrow \chi_i$, where the fractal coordinates are defined by Eq. (28). Specifically, the mapping of the fractal domain (7) with a straight cut (see Fig 14a) into the corresponding problem for the fractal continuum of the fractal dimension $D = \bar{n}\alpha$ with a straight cut of the fractal dimension $D_S = (\bar{n}-1)\alpha$, where $\bar{n}$ is the dimension of an elastic problem under consideration (see Fig 14b), suggest that the tensile stress envelope ahead of the straight cut of length $2a$ in the material with the fractal domain (7) under tensile plane stress ($\bar{n} = 2$) behave as

$$\sigma_1 = \frac{\sigma_\infty}{\sqrt{2}}\left(\frac{a}{r}\right)^{\alpha/2}\cos\left(\frac{\theta}{2}\right)\left[1 - \sin\left(\frac{\theta}{2}\right)\sin\left(\frac{3\theta}{2}\right)\right]$$

$$\tag{75}$$

$$\sigma_2 = \frac{\sigma_\infty}{\sqrt{2}}\left(\frac{a}{r}\right)^{\alpha/2}\cos\left(\frac{\theta}{2}\right)\left[1 + \sin\left(\frac{\theta}{2}\right)\sin\left(\frac{3\theta}{2}\right)\right]$$

for $r > \xi_0$ (see Fig 14c). In real fractal materials with the fractal domain (7), the stress behavior (75) is expected to be observed at distances $r > \Lambda_p >> \xi_0$, where $\Lambda_p$ is the size of plastic zone ahead of the notch tip (see Ref. [131]). Notice that Eq. (75) converts into the classical expressions of the notch tip stresses in the homogeneous (Euclidean) continuum in the limit of $\alpha = 1$, whereas in the case of fractal continuum with $\alpha < 1$, the notch tip stresses (75) are less divergent [132]. In this regard, it should be emphasized that the actual stress fields in the fractal domain (7) are essentially discontinuous non-analytic functions in $E^3$, while Eq. (75) represents analytic envelopes of the stress distributions (see Fig. 14c).

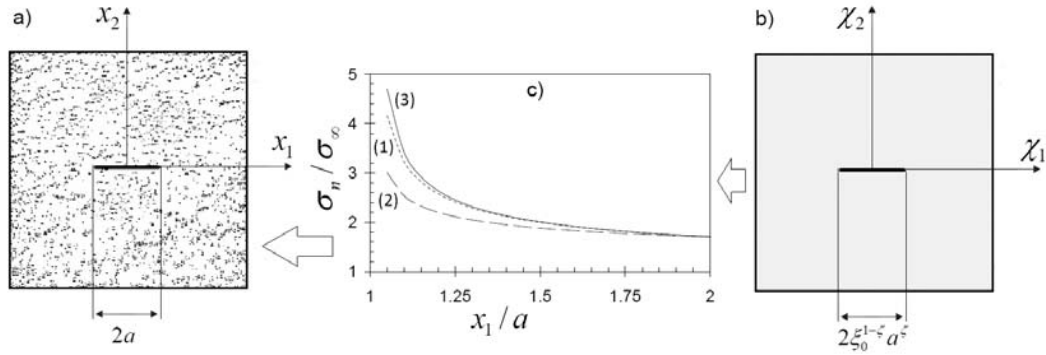

**Figure 14**. Mapping a two-dimensional elastic problem ($d = 2$) for fractal medium of $D = 3\alpha < d_\ell = 3$ with a straight cut of length $2a$ (a) into the corresponding problem for fractal continuum (b) and the overall stress distributions ($\sigma_n / \sigma_\infty$ versus $x_1 / a$) ahead of the cut tip (c) in the heterogeneous materials with: the Euclidean (1) and fractal ($D < d_\ell = 3$) (2) heterogeneities ($a / \xi_0 = 50$) and in homogeneous continuum (3). Notice that actual local stress distribution in the fractal material is essentially discontinuous non-analytic function of the Cartesian coordinates, whereas graphs in the panel (c) represent analytic envelopes of these distributions.

To analyze the stress distribution in the vicinity of a fractal crack in a heterogeneous material with linearly elastic fractal domain $\Phi_3^D \subset E^3$ having the intrinsic fractal dimension $d_\ell < 3$ (e.g., Menger sponge shown in Figs. 3 and 6), let us first consider an open region $\Omega_f \subset \Phi_3^D \subset E^3$ with a fractal boundary $\partial\Omega_f$ having the fractal dimension $D_\partial$ (see Fig. 15a). To define the trace of the displacement field $\bar{u} = (u_i) \in \Phi_3^D \subset E^3$ to the fractal boundary $\partial\Omega_f$, the fractal region $\Omega_f$ can be mapped into the fractal continuum region $\Omega \subset \Phi_D^3 \subset E^3$ with boundary $\partial\Omega$ of the fractal dimension $D_\partial$ (see Fig. 15b). Following to Panagouli [133] let us consider the classical Sobolev space $W_k^p(\Omega)$ of $L^p(\Omega)$ functions with distributional derivatives up to order $k$ in $L^p(\Omega)$, which is equipped with the p-norm. The mapping $\Omega_f(\Phi_3^D) \to \Omega(\Phi_D^3) \subset E^3$ implies that

$$k = \zeta_i \text{ and } p = 2\gamma, \tag{76}$$

while the norm and metric in the fractal continuum $\Phi_D^3 \subset E^3$ are defined by Eqs. (32) - (34). A function $\upsilon \in L^1(\Omega)$ can be define "strictly" at the point $\chi \in \Omega \cup \partial\Omega$ if the limit

$$\tilde{\upsilon}(\chi) = \lim_{r \to 0} \frac{1}{A\left[B(\chi, r_\chi) \cap \Omega\right]} \int_{B(\chi, r_\chi) \cap \Omega} \upsilon d\Omega \tag{77}$$

exists [134], where $B(\chi, r)$ is the ball in $\Phi_D^3$ with center at $\chi$ and radius $r_\chi$ and $A\left[B(\chi, r_\chi) \cap \Omega\right]$ is the area of the intersection of this ball with $\Omega$. If the limit (77) exists, then the trace of $\upsilon$ to $\partial\Omega$ can be defined as $\upsilon|_{\partial\Omega} = \tilde{\upsilon}(\chi)$ at every point $\chi \in \Omega \subset \Phi_D^3$.

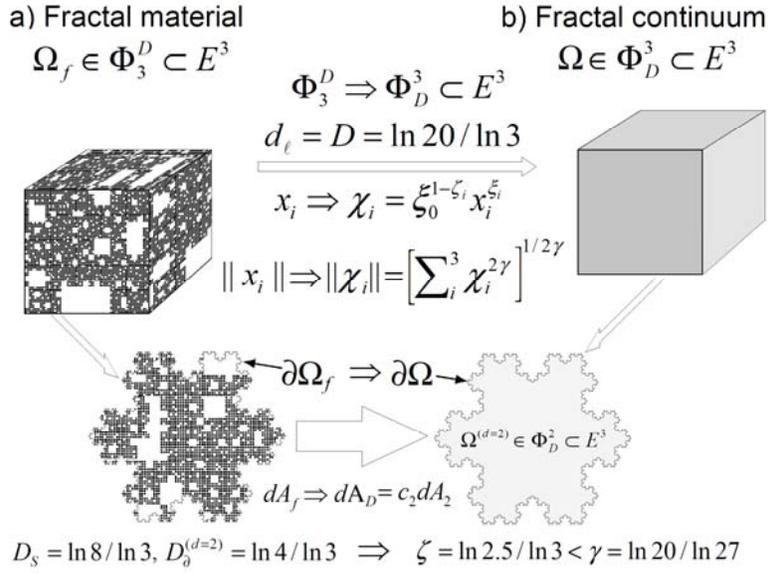

**Figure 15**. Mapping a two-dimensional elastic problem for fractal medium $\Omega_f$ with fractal boundary $\partial \Omega_f$ into the corresponding problem for fractal continuum $\Omega$ with boundary $\partial \Omega$.

In the classical linear elasticity of the homogeneous continuum ($D = n$) the displacement field on the smooth (differentiable) boundary of dimension $D_\partial = n - 1$ can be considered as an element of $H^1(\Omega)$, that is of the Sobolev space $W_1^2(\Omega)$. Consequently, the Sobolev space $H^{1/2}(\partial \Omega)$ is the space for the displacements on $\partial \Omega \in E^{n-1}$ [135]. Its dual space $H^{-1/2}(\partial \Omega)$ is the space of the boundary tractions $t_i = \sigma_{ij} n_j$. Consequently, the stresses and strains ahead of a smooth (differentiable) cut obey the asymptotic behavior $\sigma \propto \varepsilon \propto 1/\sqrt{r}$ [131]. On the fractal boundary of dimension $D_\partial > n - 1$ the displacement field $\upsilon_i \in H^1(\Omega)$ does not possess a trace $\upsilon_i \vert_{\partial \Omega}$ in $H^{1/2}(\partial \Omega)$ [136]. Nonetheless, for the case of homogeneous (Euclidean, $D = n$) continuum with the fractal boundary having the fractal dimension $n-1 \le D_S \le n$, Wallin [133] has proved that the trace operator can be defined as a bounded linear surjective operator

$$Tr : u \in W_k^p(\Omega) \to u \mid_{\partial\Omega} \in B_\varphi^{p,p}(\partial\Omega) \, , \qquad (78)$$

where $B_\varphi^{p,p}(\partial\Omega)$ is the Besov space,

$$\varphi = k - (n - D_\partial) / p \, , \qquad (79)$$

The above result implies that in the linear elastic Euclidean body ($D = n$, $k = 1$, and $p = 2$) with the fractal boundary of dimension $n-1 < D_\partial \le n$, the boundary displacement is $u_i \in B_\varphi^{2,2}$, where $\varphi = 1 - 0.5(\bar{n} - D_\partial) > 0.5$, while $\bar{n} - 1 \le D_\partial \le \bar{n}$ and $\bar{n} = 2 < n = 3$ for a plane stress problem, whereas $\bar{n} = 3$ for a three-dimensional stress problem. [22]. Consequently, the strains ($\varepsilon \propto \partial u / \partial r$) and stresses ($\sigma \propto \varepsilon$) ahead of the fractal crack in the linearly elastic Euclidean continuum (see Fig. 16a) are expected to obey the asymptotic behavior

$$\sigma \propto \varepsilon \propto K_F r^{-\varsigma} \qquad (80)$$

with the scaling exponent

$$0 \le \varsigma = (\bar{n} - D_\partial) / 2 < 0.5 \qquad (81)$$

for $a, r > \Lambda_p >> \xi_0$, where $K_F \propto \sigma_\infty a^\varsigma$ is the fractal stress intensity factor (see Fig. 16b). The local fractal dimension of self-affine crack is equal to

$$D_{\partial} = (\bar{n}-1)/H \text{ , if } H > H^* = (\bar{n}-1)/\bar{n} \text{ , whereas } D_{\partial} = \bar{n} \text{ , if } H \le H^* \text{ ,} \qquad (82)$$

where $H$ is the crack roughness (Hurst) exponent [120], Consequently, in the homogeneous materials the stress concentration exponent is equal to

$$\varsigma = \frac{\bar{n}H - (\bar{n}-1)}{2H} \text{ , if } H > H^* = (\bar{n}-1)/\bar{n} \text{ , whereas } \varsigma = 0 \text{ , if } H \le H^* \text{ ,} \qquad (83)$$

such that there is no stress concentration ahead of self-affine crack in linearly elastic Euclidean continuum, if the crack roughness exponent $H \le H^*$ [137].

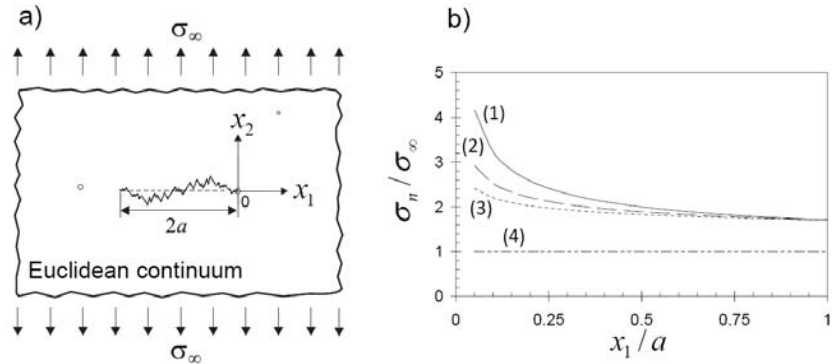

**Figure 16**. (a) Rough crack in linearly elastic heterogeneous material under tensile stress ($d = 2$) and (b) the overall stress distribution ($\sigma_n / \sigma_\infty$ versus $x_1$) ahead of a differentiable (Euclidean, $H = 1$) rough crack (1) and self-affine cracks with: $H = 0.75 > H^* = 1/2$ (2), $H = 0.75 > H^* = 1/2$ (3), and $0 < H < H^* = 1/2$ (4).

For fractal domains with $d_\ell \le n$ and the boundary of the fractal dimension $n-1 < D_{\partial} < n$ (see Fig 15) Eqs. (78) and (79) can be generalized in a straightforward manner. Namely, in the region $\Omega \subset \Phi_D^3 \subset E^3$ with the boundary $\partial\Omega$ the trace operator takes the form

$$Tr : \upsilon \in W_k^p(\Omega) \rightarrow \upsilon \mid_{\partial\Omega} \in B_\varphi^{p,p}(\partial\Omega) , \qquad (84)$$

where

$$\varphi = \zeta - \frac{D - D_\partial}{2\gamma} , \qquad (85)$$

while $\gamma = d_\ell / 3 \leq 1$ characterizes the topology of the fractal domain $\Phi_3^D$ and thus governs the metric (33) in the fractal continuum. Therefore, the stresses ahead of fractal crack in a heterogeneous material with a linearly elastic fractal domain $\Phi_3^D$ are expected to obey the power-law asymptotic behavior (80) with the scaling exponent

$$\varsigma = \frac{D - D_\partial}{2\gamma} , \text{ if } D_\partial < D , \qquad (86)$$

whereas if $D_\partial \geq D > \bar{n} - 1$, there is no stress concentration ($\varsigma = 0$) ahead of the fractal (self-affine) crack. Notice that in the case of the fractal domain (7) with a straight (plane) cut of the fractal dimension $D_\partial = (\bar{n} - 1)\alpha$, Eq. (86) predicts the same stress concentration exponent $\varsigma = 0.5\alpha$ as it is given by Eq. (75).

Furthermore, from Eq. (86) follows that the rate of stress concentration envelope in the vicinity of fractal (self-affine) crack in a material with the fractal heterogeneity can be either smaller ($\varsigma < 0.5$), if $D - D_\partial < \gamma$, or larger ($\varsigma > 0.5$), if $D - D_\partial > \gamma$, than the stress concentration rate at the tip of a smooth cut in the Euclidean continuum (see Fig. 17). Specifically, in the fractal material with a straight (plane) cut of dimension $D_S$ the cut tip stresses will be more divergent $\zeta > 0.5$ than in the classical continuum if

$$D - D_S = \zeta > \gamma \,, \tag{87}$$

as this is expected, for example, in the case of fractal domains with $d_\ell < D < 3$ obeying the Mandelbrot rule of thumb (5), e.g. the percolation clusters in 3D [53]. The stronger singularity of notch tip stresses is expected to assist the crack initiation. So, one can expect that in such materials the crack initiation stress will decrease with the cut length faster than in the Euclidean continuum. Although this prediction seems somewhat unexpected, it can be easily verified in experiments with model materials obeying the inequality (87).

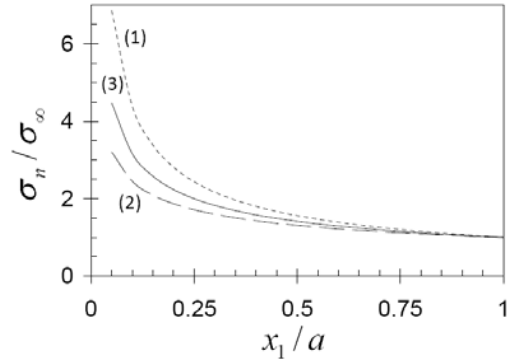

**Figure 17**. Overall stress distributions ( $\sigma_n / \sigma_\infty$ versus $x_1 / a$ ) ahead of a straight cut ( $d = 2$ ) in linearly elastic fractal materials with: $\gamma = 0.7 < \zeta = 0.9$ (1), $\gamma = 0.9 > \zeta = 0.7$ (2), and their comparison with the stress distribution ahead of a straight cut in a linearly elastic homogeneous ( $\gamma = \zeta = 1$ ) material (3).

Controversially, if $D - D_S = \zeta < \gamma$ , as this is in the case of Menger sponge shown in Fig. 6, the notch tip stress envelope $\sigma \propto a^\varsigma$ increase with the cut length more slowly than in the Euclidean continuum. This gives rise to the fractal scale effect observed in experiments with fractal materials within the range of $\xi_0 << \Lambda_p < a < \xi_C$ [37,125,138,139,140]. Furthermore, if $D - D_S = 0$, there is no stress concentration ($\varsigma = 0$) ahead of a straight cut. In this case the crack initiation stress should be independent on the cut length, as it was observed in experiments with some kinds of paper having the fractal microstructure [37]. At the same time, the roughness of a crack growing in a material with a fractal microstructure is characterized by the Hurst exponent $H = D - (d - 1)$, where $D$ is the fractal dimension of the material microstructure [127]. Therefore, the fractal dimension $D_S = (\bar{n} - 1)/H$ of the growing cracks is always less than the fractal dimension of the material microstructure. So, self-affine cracks in materials with a fractal microstructure will always propagate due to the stress concentration ahead the crack tip.

Summarizing, the fractal topology governing the norm (32) facilitate the crack initiation from the straight notch in heterogeneous material with a fractal domain of $d_\ell < 3$, whereas the fractal metric (34) obstruct the crack initiation and propagation. At the same time, the fractal (self-affine) roughness always hinders the crack propagation. Accordingly, the crack propagation in a fractal material is controlled by the interplay between its topological and metric properties and the crack roughness.

## VI. Conclusions

Fractal continuum mechanics comes into play when one studies the mechanics of heterogeneous materials with fractal domains in a smoothed picture that does not go into detail about the forces and motions of the sub-scale constituents. In this work we suggest the fractal continuum approach which allows us to account for the metric, topological, and dynamical properties of fractal domains in heterogeneous materials. The kinematics of fractal continuum deformations is developed. The Jacobian of transformations is established. The concept of stresses in the fractal continuum is defined. The mapping of mechanical problems for the fractal domains into the corresponding problems for the fractal continuum is elucidated.

In this background some specific problems are analyzed. Specifically, the stress and strain distributions in elastic fractal bars are derived. An approach to fractal bar optimization is suggested. Some noteworthy features of elastic waves in fractal materials are outlined. The effects of material metric and topology on the stress fields ahead of straight (plane) cuts and self-affine cracks in fractal materials are discussed. It is shown that the fractal nature of heterogeneity can either delay or facilitate the crack initiation and propagation, depending on the interplay between the metric and the topological properties of the fractal domain. Generalization for heterogeneous materials with elasto-plastic fractal domains can be performed in a straightforward manner. Accordingly, we expect that our findings related to the mechanics of fractal bars, elastic wave propagation, and crack mechanics in fractal materials will stimulate experimental research on these topics.

**Acknowledgements**

Author thanks Rodolfo Camacho-Velázquez and Armando Garcia Jaramillo for fruitful discussions. This work was supported by the Mexican Petroleum Company (PEMEX) under the research SENER-CONACYT Grant No. 143927.